\documentclass[journal=jpcafh,manuscript=article]{achemso}

\usepackage[utf8]{inputenc}
\usepackage[OT1]{fontenc}
\usepackage{graphicx}
\usepackage{amssymb}
\usepackage{mathrsfs}
\usepackage{amsmath}
\usepackage{xcolor}
\usepackage{latexsym}
\usepackage{amsfonts}
\usepackage{hyperref} 
\usepackage{wasysym}
\usepackage{dcolumn}
\usepackage{bm}
\usepackage[percent]{overpic}
\usepackage{natbib}
\usepackage{booktabs}
\usepackage{gensymb}
\usepackage{lineno}

\usepackage{chemformula}
\usepackage{pdfpages}
\newcommand{\ld}{\lambda_{\scriptscriptstyle D}}
\newcommand{\lb}{\lambda_{\scriptscriptstyle B}}
\newcommand{\dht}{Debye-H\"uckel }
\DeclareMathOperator\erfc{Erfc}

\title{Numerical insights on ionic microgels:\\ structure and swelling behaviour}

\author{Giovanni Del Monte}
 \affiliation{Physics Department, Sapienza University of Rome, Piazzale A. Moro 2 00185 Rome, Italy}
 \alsoaffiliation{Center for Life NanoScience, Istituto Italiano di Tecnologia, Rome, Italy}
 \alsoaffiliation{CNR-ISC, Sapienza University of Rome, Piazzale A. Moro 2 00185 Rome, Italy}
 \email{giovanni.delmonte@uniroma1.it}
 \author{Andrea Ninarello}
\affiliation{CNR-ISC, Sapienza University of Rome, Piazzale A. Moro 2 00185 Rome, Italy}
\alsoaffiliation{Physics Department, Sapienza University of Rome, Piazzale A. Moro 2 00185 Rome, Italy}%
\author{Fabrizio Camerin}
\affiliation{CNR-ISC, Sapienza University of Rome, Piazzale A. Moro 2 00185 Rome, Italy}
\alsoaffiliation{Department of Basic and Applied Sciences for Engineering, Sapienza University of Rome, via A. Scarpa 14, 00161 Rome, Italy}
\author{Lorenzo Rovigatti}
 \affiliation{Physics Department, Sapienza University of Rome, Piazzale A. Moro 2 00185 Rome, Italy}
 \alsoaffiliation{CNR-ISC, Sapienza University of Rome, Piazzale A. Moro 2 00185 Rome, Italy}
 \author{Nicoletta Gnan}
\affiliation{CNR-ISC, Sapienza University of Rome, Piazzale A. Moro 2 00185 Rome, Italy}
 \alsoaffiliation{Physics Department, Sapienza University of Rome, Piazzale A. Moro 2 00185 Rome, Italy}
\author{Emanuela Zaccarelli}
\affiliation{CNR-ISC, Sapienza University of Rome, Piazzale A. Moro 2 00185 Rome, Italy}
 \alsoaffiliation{Physics Department, Sapienza University of Rome, Piazzale A. Moro 2 00185 Rome, Italy}%
 \email{emanuela.zaccarelli@cnr.it}

\begin{document}
\begin{abstract}
Recent progress has been made in the numerical modelling of neutral microgel particles with a realistic, disordered structure. In this work we extend this approach to the case of co-polymerised microgels where a thermoresponsive polymer is mixed with acidic groups. We compare the cases where counterions directly interact with microgel charges or are modelled implicitly through a \dht description. We do so by performing extensive numerical simulations of single microgels across the volume phase transition varying the temperature and the fraction of charged monomers. We find that the presence of charges considerably alters the microgel structure, quantified by the monomer density profiles and by the form factors of the microgels, particularly close to the volume phase transition (VPT).  We observe significant deviations between the implicit and explicit models, with the latter comparing more favourably to available experiments. In particular, we observe a shift of the VPT temperature to larger values as the amount of charged monomers increases. We also find that below the VPT the microgel-counterion complex is almost neutral, while it develops a net charge above the VPT. Interestingly, under these conditions the collapsed microgel still retains a large amount of counterions inside its structure. Since these interesting features cannot be captured by the implicit model, our results show that it is crucial to explicitly include the counterions in order to realistically model ionic thermoresponsive microgels.
\end{abstract}

%\makeatletter
%\setlength\acs@tocentry@height{8.3cm}
%\setlength\acs@tocentry@width{4.5cm}
%\makeatother
%\begin{tocentry}
%\includegraphics{toc_}
%\end{tocentry}

\maketitle

\section{\label{sec:introduction}Introduction}
Microgels are colloidal-scale polymer networks which have recently become a favourite model system\cite{yunker2014physics,fnieves_lyon,brijitta2019responsive,karg2019nanogels} thanks to their intrinsic softness and to the possibility to respond to external stimuli with changes in size. Such a phenomenon is commonly referred to as Volume Phase Transition (VPT)~\cite{fernandez2011microgel} and it is controlled by the properties of the constituent polymers.  The prototype example is given by Poly(N-isopropyl-acrylamide), PNIPAM, a thermoresponsive polymer which gives microgels the ability to reversibly increase or reduce their size after a change of temperature around the so-called VPT temperature $T_{\rm VPT} \sim 32^{\circ}$ C. 
Another interesting case can be realized using pH-responsive ionic polymers, made of weak
acidic or weak alkaline monomers. The resulting ionic (or simply called charged) microgels are able to adjust their bare charge in response to a pH variation by releasing \ch{H+} or \ch{OH-} ions due to the dissociation of a fraction of monomers~\cite{synthesis}.

Out of the many possibilities provided by modern-day synthesis methods, co-polymerised PNIPAM-co-PAAc microgels are of particular interest~\cite{synthesis,ionic_schurt,seiffert,fnieves_multiresponsive}, as they combine  the thermoresponsive properties of PNIPAM with the pH-responsive features of polyacrylic acid (PAAc), stemming from the weak acidic nature of AAc monomers. Indeed, at low pH almost all AAc monomers are not dissociated because of the high concentration of \ch{H^+}, which favours the inverse recombination reaction that leads to an almost neutral network. On the other hand, for high pH values, most of the acidic monomers dissociate, generating a charge distribution throughout the particle volume.  It is important to note that the fraction and distribution of these charges within the network depend on the chosen experimental conditions, such as the packing fraction, the specific molecular interactions, the local counterions concentration and the electrostatic interactions between nearest charged monomers, which can be optionally mediated by the presence of salt\cite{fernandez2011microgel,fnieves_multiresponsive,truzzolillo_sennato}. 

The multiresponsive character of ionic microgels makes them highly versatile. They are indeed 
responsive also to external alternating electric fields, through which their mutual interactions (and hence their phase behaviour) can be tuned\cite{ionic_selfassembly_1, ionic_selfassembly_2}. Their single-particle properties have been extensively investigated in experiments as a function of both temperature and pH\cite{fnieves_multiresponsive}. Microgels with different content of AAc obtained through several synthesis methods have been analysed in order to assess the effects of inhomogeneities in the distribution of crosslinkers and charged monomers\cite{seiffert,ionic_schurt}. The tunability of ionic microgels has also been exploited in several fields of research, from biology\cite{uptake_release} to laser technology\cite{photonic_crystal}. For instance, their dual responsiveness makes them highly suitable to be employed in the smart design of switch optical devices \cite{fnieves_lyon} based on colloidal photonic crystals. 

\noindent
Understanding the effects of electrostatic interactions in ionic microgels could also shed light on the behaviour of other kinds of microgels. Indeed, even those constituted by PNIPAM only (often considered as neutral microgels) show interesting features, particularly above the VPT temperature, which suggest the presence of charge effects\cite{howe,truzzolillo_sennato}. In addition, microgels consisting of two different interpenetrated networks, made of PNIPAM and PAAc respectively, have recently gathered a lot of attention because of their suitability to study the problem of fragility in structural glasses\cite{mattsson,valentina}.

From the theoretical point of view, several investigations of the swelling of charged microgels, mostly relying on a mean-field treatment of the polymer network based on the Flory-Rehner theory\cite{degennes}, have been reported. In these works, electrostatic effects and steric interactions due to the presence of counterions have been taken into account by approximated theories such as the Poisson-Boltzmann equation\cite{denton2}, the Ornstein-Zernike integral equation\cite{moncho_OZ} and density functional theory\cite{moncho_DFT}.
Also an effective interaction potential has been derived using linear response theory\cite{denton1}, which made it possible to draw a phase diagram for ionic microgels\cite{likos_mgel_book}.
On the numerical side, the use of coarse-grained models\cite{ionic_selfassembly_2,denton_swelling_suspension} has allowed to go beyond the mean-field framework and tackle the behavior of ionic microgels at all concentrations. However, in order to refine the highly coarse-grained models required to study the bulk properties of microgel suspensions, it is important to first correctly capture the single-particle behavior, a task that has been tackled only relatively recently\cite{rovigatti_review, quesada_review} due to the high computational cost of numerical studies reproducing microgels at monomer-resolved level.

The inclusion of long-range electrostatic interactions on complex objects such as microgels is a challenging and numerically demanding task, particularly if counterions are explicitly considered. Therefore, in several cases, an implicit treatment of counterions, for example based on the \dht theory, has been employed to make it possible to perform simulations of relatively large systems~\cite{winklerDH,kobayashi_salt}. 
However, a few numerical investigations have also been carried out in the explicit presence of the counterions. A pioneering work reported coarse-grained simulations of polyelectrolyte gel networks~\cite{linse_hydrogel}, while simulations of single nanogel particles have appeared only later on~\cite{holm_cions,quesada2012cions,quesada2013cions,jha2011theoretical,schroeder}. 
Several techniques have been devised to treat charged networks. Particularly, recent Monte Carlo simulations\cite{MC_aachen,reaxchain,effective_int} have been carried out to provide a coarse-grained description of the dissociation reaction on a statistical basis. These studies concluded that all investigated macroscopic properties mostly depend on the number of charges, rather than on their distribution, in agreement with experimental observations\cite{seiffert}. Notwithstanding this, all coarse-grained studies of ionic microgels have so far been performed with networks built out of ordered topologies, \textit{e.g.} based on the diamond lattice, which cannot take into account the disordered nature of real polymer networks\cite{rovigatti_review}.

In order to go beyond mean-field and to account in a more realistic way for the effect of the network topology, in this work we perform extensive simulations of charged microgels modelled as disordered networks. We start by preparing neutral microgel configurations following previous works\cite{silicomicrogel,andrea2019preprint}, ensuring that the internal microgel structure reproduces the swelling behavior and form factors of experimental non-ionic microgels. Then, we add a quenched charge distribution, varying the fraction of charged monomers that are randomly distributed  throughout the network. Since the probability that a monomer is charged is lower near crosslinkers\cite{MC_aachen,reaxchain}, we add the constraint that the latter are always neutral. To account for charge-charge interactions we perform two different kinds of simulations: (i) we rely on the \dht model in which charged monomers interact implicitly through a two-body Yukawa potential and (ii) we explicitly include counterions as charged coarse-grained particles. We calculate the density profiles and form factors of the microgels for both approaches and average over different charge realizations. We simulate microgels in swollen conditions and across the volume phase transition by using a solvophobic interaction between the monomers that models the different quality of the solvent as temperature varies\cite{silicomicrogel,andrea2019preprint}.

Our work is important to understand the effects that inhomogeneous topologies and charge distributions beyond mean-field can have on the single-particle behavior of ionic microgels, filling a gap in the current literature. In addition, we provide significant insights on the difference between neutral and charged microgels across the volume phase transition. Indeed, the competition between the electrostatic repulsion and the solvophobic attraction, which develops at intermediate temperatures in between the swollen and collapsed regimes, could be important for the arising of a distinct phenomenology in the presence of charges. Finally our work can be considered as a starting point for future investigations at finite concentrations, shedding light on the deswelling behavior of ionic microgels, which, differently from neutral ones, already takes place at concentrations below the overlap one\cite{ionic_schurt,denton_swelling_suspension}.

\section{Models and methods}\label{sec:models}
\small
\subsection{Monomer interactions}
To analyse the role of charges on the single-particle properties and on the swelling behaviour of microgels, we exploit a recently proposed numerical protocol\cite{silicomicrogel} to generate disordered, heterogeneous microgels that are structurally similar to real neutral ones. We start by preparing fully connected spherically shaped networks by confining patchy particles in a cavity with a designing force on the crosslinkers, which provides the typical core-corona structure of realistic microgels\cite{andrea2019preprint}. Once the network is formed, we freeze the topology of the network and adopt a monomer-resolved approach\cite{grest1986molecular}. The beads that make up the polymers interact via a steric repusion, modeled with the Weeks-Chandler-Anderson (WCA) potential:
\begin{equation}
\label{eq:wca}
V_{\text{WCA}}(r)  =  
\begin{cases}
4\epsilon\left[\left(\frac{\sigma}{r}\right)^{12}-\left(\frac{\sigma}{r}\right)^6\right]+\epsilon & \quad \text{if} \quad r \le 2^{1/6}\sigma  \\
0 & \quad \text{if}  \quad  r > 2^{1/6}\sigma
\end{cases}
\end{equation}
where $\epsilon$ and $\sigma$ are respectively the energy and length units.
In addition, connected beads interact via the finitely extensible nonlinear elastic potential (FENE):
\begin{equation}
V_{\text{FENE}}(r)  = 
-\epsilon k_F{l_0}^2\log\left[1-{\left(\frac{r}{l_0\sigma}\right)}^2\right], \quad r < l_0\sigma 
\end{equation}
where $l_0$ sets the maximum bond distance and $k_F$ is a stiffness parameter influencing the rigidity of the bond and the equilibrium bond-distance. This potential ensures that no covalent bonds between the monomers can be broken during the course of the simulations.
In all cases, we use $k_F=15$ and $l_0=1.5$.

All monomers also interact with each other by means of an effective solvophobic potential, named $V_{\alpha}$, which implicitly takes into account the monomer-solvent interactions\cite{amphiphilic}:
\begin{equation}\label{eq:valpha}
V_{\alpha}(r)  =  
\begin{cases}
-\epsilon\alpha & \quad \text{if} \quad r \le 2^{1/6}\sigma  \\
\frac{1}{2}\alpha\epsilon\left\{\cos\left[\gamma{\left(\frac{r}{\sigma}\right)}^2+\beta\right]-1\right\} & \quad \text{if}  \quad  2^{1/6}\sigma < r \le l_0\sigma  \\
0 & \quad \text{if}  \quad  r > l_0\sigma
\end{cases}
\end{equation}
with $ \gamma = \pi \left(\frac{9}{4}-2^{1/3}\right)^{-1} $ and $\beta = 2\pi - \frac{9}{4}\gamma$\cite{amphiphilic}. This potential represents an effective attraction, modulated by the solvophobic parameter $\alpha$, arising between thermo-responsive monomers at high temperatures. In other words, $\alpha$ plays the role of an effective temperature: $\alpha=0$ represents good solvent conditions, while with increasing $\alpha$ the quality of the solvent worsens, leading to the aggregation of beads and to the shrinking of microgel particles.

We complement this model by adding electrostatic interactions between charges that are randomly assigned to a fraction $f$ of the microgel monomers. This choice aims to model the dissociation of weak electrolyte groups, usually giving rise to negatively charged microgels, such as when acrylic acid is used as a co-monomer in the synthesis process.
The neutrality of the overall suspension imposes the presence of positively charged counterions, which balances the total charge of the microgel-counterion complex. 
In the simplest approach, the effect of charges can be taken into account by using the \dht potential, which models the charge-charge interaction as a screened Coulomb (or Yukawa) potential acting between each pair of charged beads as \cite{hunter}:
\begin{eqnarray}
V_{DH}(r) &=& k_{\scriptscriptstyle B} T\frac{\lb}{r} \exp\left(-\frac{r}{\ld}\right),
\label{eq:DH}
\end{eqnarray}
where $\lb$ and $\ld$ are the Bjerrum and the Debye lengths, respectively. The former represents the distance at which two ions of valence $z$ feel a repulsive energy exactly equal to $k_{\scriptscriptstyle B} T$, thus quantifying the relative intensity of the electrostatic forces, and it is defined as:
\begin{equation}
\lb = \frac{z^2e^2}{4\pi\epsilon_0\epsilon_r k_{\scriptscriptstyle B} T} %= \frac{{q^*}^2}{T^*}\sigma 
\label{eq:bjerrum}
\end{equation}
where $\epsilon_0$ and $\epsilon_r$ are the vacuum and relative dielectric constants and $e$ is the elementary unit charge. The Debye length instead is the screening length, depending on both $\lb$ and on the density of counterions $\rho_{ci}$ as:
\begin{equation}
\ld = \left(4\pi\lb \rho_{ci}\right)^{-1/2}.
\label{eq:debye}
\end{equation}
The \dht approach can be used in principle only for symmetric electrolytes, i.e. when the valence of positive and negative ions is the same, as it is for the present case \cite{levin} since the multiple dissociation of single polyelectrolyte monomers is highly unlikely. 
We work with reduced units, with $\sigma$, $m$, $\epsilon$ being the units of length, mass and energy, respectively.
Within this unit system, the experimental Bjerrum length, that is $\lb\approx 0.7 nm$ for monovalent ions in water at room temperature, translates into a reduced Bjerrum length $\lb^*\approx 1$, assuming $\sigma \sim 1.0$~nm comparable to the Kuhn length of both neutral NIPAM and charged AAc monomers. This can be considered as a lower estimate of $\sigma$, according to different types of measurements for linear PNIPAM chains\cite{kuhn_richtering}. Note that the use of a larger value of $\sigma$ would significantly decrease $\lb$, thus resulting in a very small effect of the \dht repulsion as compared to the neutral case.

Although the \dht model is suitable to  implicitly treat the role played by counterions in homogeneous systems and in the effective interactions among colloidal particles in dispersions, it should be avoided when studying electrostatic ion-ion interactions within inhomogeneous weak-electrolyte systems such as charged polymeric particles. In particular, one of the drawbacks of using this approach is that we cannot easily compare the gyration radius as a function of $\ld$ with the experimentally measured diameter of the microgel varying the pH of the suspension or the salt concentration. Indeed, for weak poly-electrolytes, there is not a simple link between the pH and the dissociation fraction of the acidic monomers, which determines the value of $\ld$ \cite{weak_ion_simulations}. Moreover, this model cannot take into account other relevant effects due to the presence of counterions, such as their osmotic pressure.
In order to overcome these issues it is crucial to explicitly take into account the counterions and thereby to adopt an alternative model where all charged beads interact via the bare Coulomb potential, as:
\begin{eqnarray}
V_{\rm coul}(r) &=& \frac{z^2 e^2}{4\pi\epsilon_0\epsilon_r r}.
\end{eqnarray}
For ion-ion interactions this term is complemented by a steric repulsion, modeled again with the WCA potential (Eq.~\ref{eq:wca}).
This second approach significantly increases the computational cost of the simulations, but at the same time it yields a realistic representation of the counterion distributions within the network, which is important to correctly describe the behavior of the microgels across the volume phase transition.
This type of study calls for some preliminary investigations, that are described in detail in the Supporting Information. In particular, we analyzed the dependence of our results on the choice of the simulation box (see section S1), discovering that there is a critical size of the box below which the long-range electrostatic forces are not correctly taken into account. In addition, we explored the role of the counterions diameter $\sigma_c$ on the microgel swelling behavior (see section S2), finding that the use of too large counterions yields unrealistic excluded volume effects in the collapsed state of the microgel. We thus fix $\sigma_c=0.1\sigma$ throughout the rest of the manuscript.

\subsection{Numerical simulations}
We perform Molecular Dynamics simulations of single microgels with $N\sim 42000$ monomers at fixed crosslinker concentration $c=5\%$. Microgels are assembled as in Ref.~\cite{andrea2019preprint} in a spherical cavity of radius $R_0=50\sigma$, yielding an internal structure of the microgels which compares very well with experimental ones obtained through radical polymerisation techniques at the same value of $c$. Once a fully connected network is assembled, we randomly assign a charge $ze=1$ to a fraction $f$ of the monomers, maintaining this charge distribution fixed throughout the simulation run. We average results over four different topologies and three different charge distributions.

We study microgels for three different charge fractions, $f=0.05$, $0.20$, $0.95$, and for several values of the solvophobic parameter $\alpha$ across the VPT. The equations of motion of the system are integrated via the velocity-Verlet algorithm \cite{tuckerman}. The equilibration of the system is carried out in the canonical ensemble using the Nos\`e-Hoover chains thermostat for $1.6\times 10^6$ simulation timesteps, while a long production run in the microcanonical ensemble of  $\sim 2\times 10^6$ steps is used to obtain equilibrium averages of the thermodynamic observables under investigation. We used a cut-off of $R_{\scriptscriptstyle cut} = 5\ld$ for the \dht potential, whereas the long-range Coulomb interactions are computed with the particle-particle-particle-mesh method \cite{p3m}. For the latter type of simulations we used the LAMMPS package\cite{LAMMPS}.

\subsection{Main Observables} 
To assess the microgel size, we calculate the radius of gyration, defined as:
\begin{equation}
R_g = \left(\frac{\sum_{i=1}^{N}(\vec{r}_i-\vec{r}_{CM})^2 }{ N } \right)^{1/2}
\end{equation}
where $\vec{r}_i$ and  $\vec{r}_{CM}$ are the positions of the $i$-th monomer and of the microgel's center of mass, respectively.

To gain a better knowledge of the inner structure of the microgel we calculate its density profile, defined as the average density at a fixed distance from the center of mass:
\begin{equation}
\rho(r)= \left\langle \frac{\sum_{i_{=1}}^{N} \delta (|\vec{r}_{i}-\vec{r}_{CM}|-r)}{N} \right\rangle.
\label{eq:rho}
\end{equation}
We also compute the density profile of charged monomers, labelled as $\rho_{CH}(r)$, and that of counterions only, labelled as $\rho_{CI}(r)$.
By adding the two latter quantities, weighted by the respective charge, we obtain the net-charge density profile $\rho_Q(r) = -\rho_{CH}(r)+\rho_{CI}(r)$, which provides information on the charge distribution throughout the volume of the particle.

The counterpart of the density profile in Fourier space is the form factor $P(q)$, which can be readily obtained in neutron or x-ray scattering experiments of dilute microgel suspensions. In simulations $P(q)$ can be directly calculated as:
\begin{equation}
P(\vec{q}) = \left \langle \frac{1}{N}\sum_{i=1}^N\sum_{j=1}^N \exp \left[i\vec{q}\cdot (\vec{r}_{i}-\vec{r}_{j})\right] \right \rangle,
\end{equation}
where the brackets $\langle \cdot \rangle$ indicate ensemble averages and $r_{i}$ is the position of the $i$-th monomer. %\fc{lasciamo la frase che segue? troppo tecnico?} 
We have computed the rotationally invariant quantity $P(q)$ as an average of $P(\vec{q})$ over 300 vectors $\vec{q}$ randomly picked onto a spherical surface of radius $q$.

Usually, experimental and numerical data of $P(q)$ for neutral microgels are described by the fuzzy sphere model\cite{fuzzy}, which is able to account for particles with a homogeneously dense core and a fuzzy corona, wherein the density gradually decreases away from the center of mass. This results in a density profile $\rho(r)\propto\erfc\left(\frac{r-R}{\sqrt{2}\sigma_{\rm surf}}\right)$, where $\erfc(\cdot)$ is the complementary error function, while $R$ and $\sigma_{\rm surf}$ are related to the extension of the core and of the corona, respectively.
 However, it has recently been shown by super-resolution microscopy measurements~\cite{superresolution} that the assumption of a homogeneous core is not accurate, and that the density profile of microgels is better approximated by the function $\rho(r)\propto\erfc\left(\frac{r-R}{\sqrt{2}\sigma_{\rm surf}}\right)(1-sr)$, which includes a linear growth of the monomer density inside the core modulated by the parameter $s$. Our microgels have thus been assembled through a numerical protocol that is able to reproduce such features~\cite{andrea2019preprint}. The additional linear term in the density profiles modifies the shape of the form factor in an extended fuzzy sphere model:
\begin{eqnarray}\label{eq:mod_fuzzy}
	P(q) &\propto& \left\{\left[ \frac{3\left(\sin(qR)-qR\cos(qR)\right)}{\left(qR\right)^3} + s\left( \frac{\cos(qR)}{q^2R} - \frac{2\sin(qR)}{q^3R^2} - \frac{\cos(qR)-1}{q^4R^3} \right)\right] \right. \nonumber\\ &\times& \left. \exp\left(-\frac{\left(q\sigma_{\mathtt{surf}}\right)^2}{2}\right) \right\}^2.
	\end{eqnarray}
This functional form is usually added to a Lorentzian term which takes into account the inhomogeneities of the network at large $q$. However, such a term was often found to be unsatisfactory in comparison to available experiments, especially for hydrogels\cite{tanaka_ffactors}. A step forward is represented by the modified Lorentzian proposed by Shibayama and Tanaka\cite{tanaka_ffactors}, which relies on the assumption that the spatial correlations of the network decay according to $r^{D-d}$, where $d$ is the system physical dimension and $D$ is the fractal dimension of the correlated domains. For large $q$ the form factor can thus be written as:
\begin{equation} 
\label{eq:Lorentzian}
P(q) \propto \frac{1}{\left[1+\frac{D+1}{3}\xi^2q^2\right]^{D/2}}
\end{equation}
with $\xi$ being the length over which concentration fluctuations are spatially correlated.

\normalsize
\section{Results and Discussion}
\label{sec:results}

\subsection{Swollen microgels}
\label{subsec:swollen}
In this section we discuss the properties of microgels in good solvent conditions. In our model this corresponds to $\alpha=0$, \textit{i.e.} to monomers that interact \textit{via} the bead-spring model plus the charges contribution only. To quantify the latter, we analyze both the \dht approach and the simulations in the presence of explicit counterions, carrying out a comparison between these approaches and the neutral case.

\subsubsection{\dht microgels}
\begin{figure}[!ht]
	\includegraphics[width=0.6\textwidth]{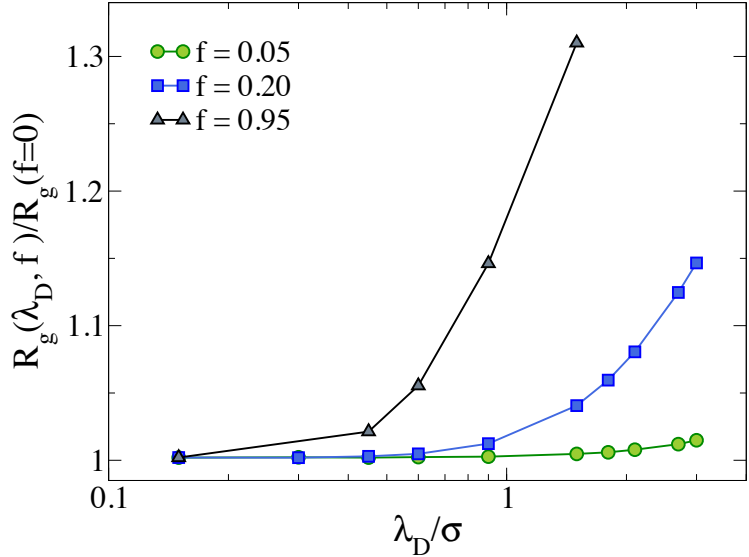}
	\caption{\small Gyration radius $R_g$ as a function of Debye length $\ld$ for different values of $f$, up to the fully charged case ($f=0.95$). Data are normalized with respect to the neutral microgel $(f=0)$.}
	\label{fig:chrnd_RgvsLd}
\end{figure}

We start by reporting in Fig. \ref{fig:chrnd_RgvsLd} the microgel radius of gyration $R_g$ for  the \dht model  as a function of the screening length $\ld$ for three different values of $f$. Data are normalized with respect to the neutral microgel case, for which $f=0$. For all considered values of $f$, the microgel size increases with $\ld$. 
We observe a progressive increase of the microgel  size as $f$ increases, with the fully charged microgel, which corresponds to $f=0.95$ since crosslinkers are not charged, displaying the strongest variation of $R_g$ with respect to the uncharged case. The fully charged situation was also analyzed in Refs.\cite{winklerDH, kobayashi_salt, MC_aachen, reaxchain, effective_int} for a diamond-like microgel and
we find comparable variation of the microgel size to that reported in these works. 

\begin{figure}[!ht]
	\includegraphics[width=0.6\textwidth]{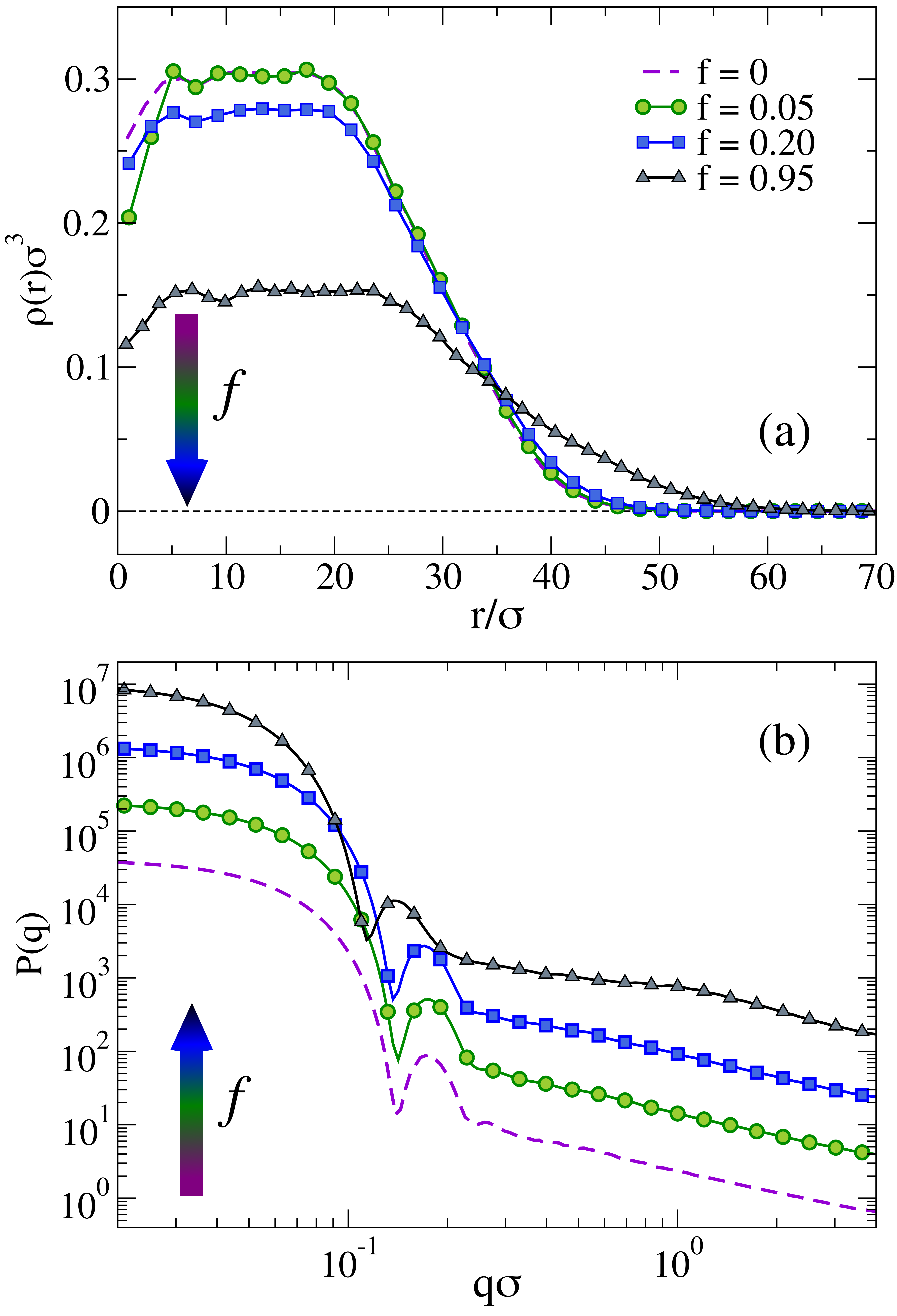}
	\caption{\small (a) Monomer density profiles and (b) form factors as a function of $f$ at fixed $\ld = 1.5\sigma$ for the swollen ($\alpha=0$) microgel, from neutral ($f=0$) to fully charged ($f=0.95)$ conditions. In (b) data are shifted on the vertical axis by a factor of $3$ with respect to each other to improve visualization.}
	\label{fig:ld} 
\end{figure}

To visualize the effect of charges on the internal structure of the microgels, we report in Fig.~\ref{fig:ld} the density profiles and the form factors of the microgels for a representative value of $\ld$ and different values of $f$, from the neutral case up to the fully charged one. As expected, we find that a larger presence of charges has the effect to lower the density of monomers in the core region and consequently to increase it in the corona, as shown in Fig.~\ref{fig:ld}(a). Such a variation of the density profile is barely noticeable for $f=0.05$ and very moderate for $f=0.20$. However, the fully charged case displays a considerably different profile, where the core density is about half of that in the neutral case and the corona extends to distances larger by about 50\% with respect to the neutral case.  Small oscillations at short distances $r$ disappear when averaging over a larger number of realizations of network topologies\cite{silicomicrogel,camerin2018modelling}.

Corresponding $P(q)$ are reported in fig.~\ref{fig:ld}(b) showing again tiny changes from $f=0$ to $f=0.20$: the first peak slightly shifts to smaller wavevectors, reflecting the larger size of the microgel, but no additional peaks are observed. In addition, the slope of the curves at high $q$ remains the same. The case $f=0.95$ shows the same features, but amplified by the larger number of charges.
Interestingly, we can compare the results in Fig.~\ref{fig:ld}, with those reported in Ref.\cite{winklerDH} for a fully charged diamond lattice network where charges are also modelled by a \dht potential. In that work, regular oscillations in the density profiles were observed, due to the underlying presence of a regular mesh of the network, as also discussed previously for non-ionic microgels\cite{rovigatti_review}. Such oscillations were further enhanced in the presence of charges, leading to unrealistic density profiles. Similarly, the form factors were found to display strong deviations from the fuzzy sphere model, displaying a minimum at intermediate $q$. Such features are totally absent in the disordered network model examined here, regardless of the amount of charge. These results confirm once more the importance of a correct modeling of the underlying network topology to treat single-particle microgel properties, also for charged microgels.

We notice that at high pH the average fraction of ions that dissociate from the microgels may be considerably lower than the ideal one for dilute suspensions of AAc, resulting in a larger average distance between charged monomers\cite{weak_ion_simulations}. This poses concerns about the use of too large values of $f$, which would be unrealistic under these conditions.  Indeed, if we look more carefully, we notice that $P(q)$ for $f=0.95$ displays a sort of kink for $q\sigma\sim 1$. Looking at the snapshots of the corresponding microgel (not shown), evident holes appear in the structure with a size comparable to this length scale, suggesting that such high-charge conditions are probably far from realistic ones for standard co-polymerized microgels. For these reasons, in the following, we will consider only the $f=0.05$ and $f=0.20$ cases.

\begin{figure}[!ht]
	\includegraphics[width=0.6\textwidth]{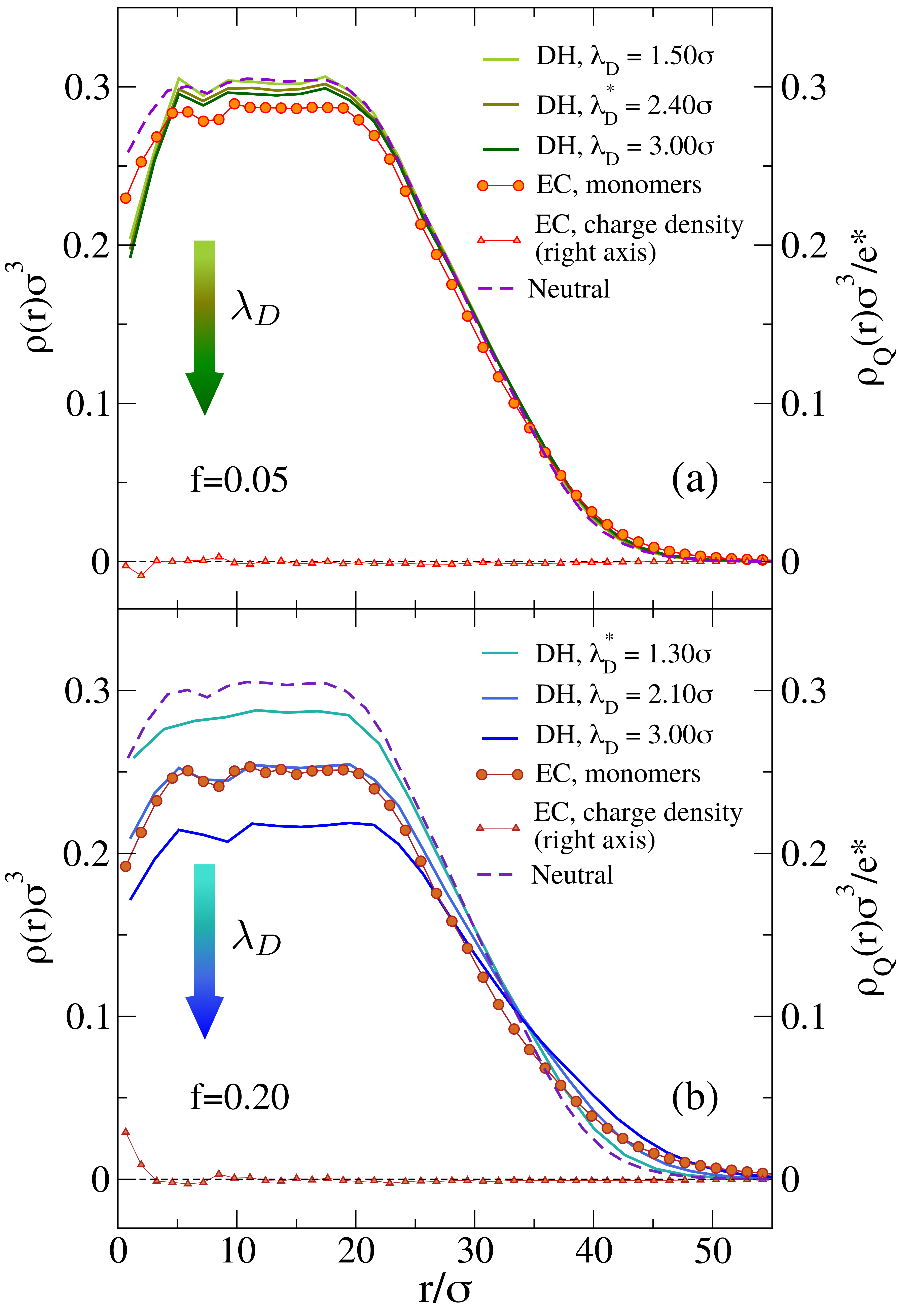}
	\caption{\small Evolution of the monomer density profiles for the \dht microgels (DH) with different $\ld$ (full lines) and for the model with explicit counterions (EC, solid circles) with (a) $f=0.05$ and (b) $f=0.20$, in the swollen state ($\alpha=0$). The net charge density profile $\rho_Q(r)$ for the EC microgel (solid diamonds) is also reported (scale on the right axis). The density profiles of the corresponding neutral microgel (dashed lines) are shown for comparison.}
\label{fig:chrnd_prof} 
\end{figure}

\subsubsection{Microgels with explicit counterions}

We extend our study to the explicit counterions (EC) model by focusing on two values of $f=0.05, 0.20$. Fig.~\ref{fig:chrnd_prof} reports the resulting density profiles comparing the explicit model results to the \dht ones (DH) for different values of $\ld$.

For $f=0.05$ the two models yield similar results, probably due to the limited presence of charged monomers. However, for $f=0.20$ the microgel with explicit counterions exhibits a more extended corona and a less dense core than the \dht model for all investigated values of the Debye length. Even a large increase of $\ld$, which has a qualitatively similar effect to the increase of $f$ (since we find fewer monomers in the core and a more extended corona), gives rise to results that do not superimpose onto the explicit counterions case, suggesting an intrinsic different structure of the microgels between the two models.
In an attempt to set up an effective \dht model that mimics the explicit one, we have calculated an effective screening length $\ld^*$ from Eq.~\ref{eq:debye} by substituting $\rho_{ci}$ with the average density of counterions that is present inside the microgel with explicit counterions within a sphere of radius $(2/3)R_g$. Such a value roughly takes into account the whole extent of the core region.
In this way, we obtain $\ld^*\simeq2.4\sigma$ for $f=0.05$ and $\ld^*\simeq1.3\sigma$ for $f=0.2$, respectively. 
The resulting density profiles of the $\ld^*$-microgels are reported in Fig.~\ref{fig:chrnd_prof}, showing also in this case a different behavior with respect to the explicit model. Deviations are larger in  Fig.~\ref{fig:chrnd_prof}(b) for the higher fraction of charges considered, where the effective \dht result is actually much more similar to the neutral case than to the explicit one.

The use of explicit counterions  makes it possible to monitor  the total charge density of the microgels $\rho_Q(r)$, also shown in Fig.~\ref{fig:chrnd_prof}. For both values of $f$ we find that the complex microgel-counterions is globally neutral at all length scales. Indeed, charge density profiles are much smaller with respect to the average inner densities of charged monomers both for $f=0.05$ ($\rho_{Q}\sim 1.5\cdot10^{-2}\sigma^{-3}$) and $f=0.20$ ($\rho_{Q}\sim 5.0\cdot10^{-2}\sigma^{-3}$).
Thus, the counterions are able to freely diffuse throughout the microgels, even within the core, so that they fully counteract the electrostatic repulsion. Tthe presence of the counterions inside the network thus contributes to the increase of the size of the microgel.

\begin{figure}[!ht]
	\includegraphics[width=0.6\textwidth]{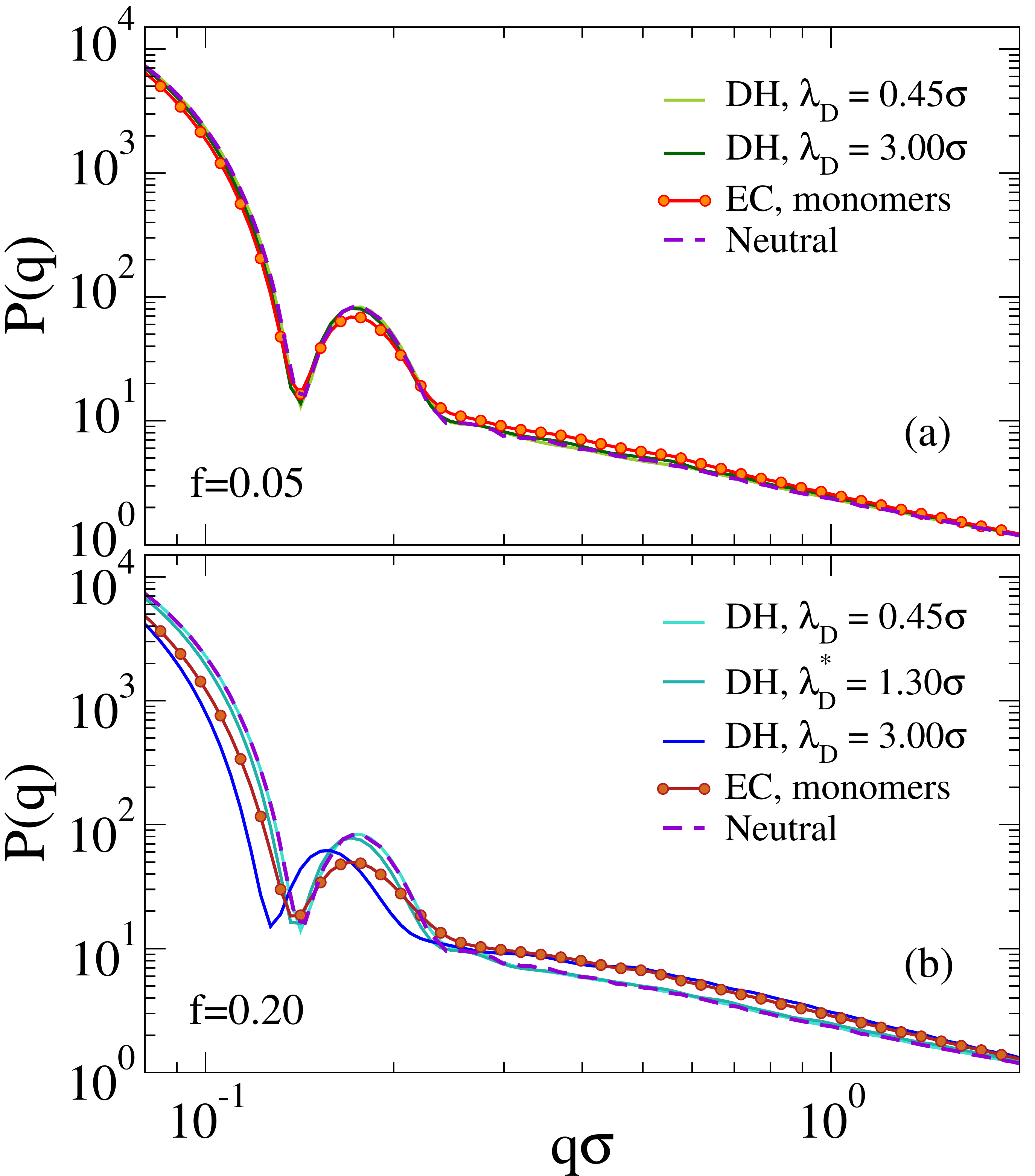}
	\caption{\small Form factors for the \dht microgels (DH) by varying $\ld$ (solid lines) and for the model with explicit counterions (EC, solid circles) with (a) $f=0.05$ and (b) $f=0.20$ in the swollen state ($\alpha=0$). The form factors for the corresponding neutral microgel (dashed lines) are reported for the sake of comparison. }
	\label{fig:chrnd_ff} 
\end{figure}

The behavior of the form factors is shown in Fig.~\ref{fig:chrnd_ff}. We start by discussing the results for $f=0.05$ in Fig.~\ref{fig:chrnd_ff}(a), where only very minor changes to $P(q)$ are observed and no shift of the first peak position is found. We find that 
all curves corresponding to the \dht model are quite similar to the neutral case, independently of $\ld$. The only noticeable difference is a weakening of secondary peaks in the presence of charges. The explicit model is the only one with a significantly smaller peak height and a different behavior at larger $q$, with some small residual oscillations and an apparently different slope at intermediate wavevectors.

These features are amplified for $f=0.20$, where now also a shift of the first peak position to smaller $q$ values is observed. This is actually more evident for the implicit, rather than for the explicit model, which displays the smallest peak intensity. Again, secondary peaks are suppressed and now the appearance of a different slope for $P(q)$ in the second peak region is more evident.  Hence, we confirm that the \dht model cannot be superimposed on the one with explicit counterions, even with the use of an effective \dht model with $\ld=\ld^*$.

The fact that the implicit \dht model fails to reproduce the features observed in the explicit counterions case can be attributed to at least two reasons. First, the permeable and inhomogeneous structure of microgels as well as the presence of a rough interface among its inner part and the solvent generate uneven distributions of charges. These in turn lead to different screening conditions in different regions of the particle, that cannot be captured by the single lengthscale of the \dht model. Second, the counterions have to balance the electrostatic attraction which drives them close to the charged monomers of the network, and the entropic gain that pushes them to leave the microgel, the latter being particularly strong for small-sized nanogels~\cite{conformational_prop_nanogels}. Under these conditions, it is not a priori trivial to assess the relative contributions to the swelling of the electrostatic interactions and of the counterions osmotic pressure, respectively.
In addition, these considerations make such a kind of implicit treatment not readily applicable to the study of finite-concentration suspensions (beyond the scope of this paper), because of the complex dependence, in thermosensitive soft colloids, of the local counterions concentration on the effective packing fraction and on temperature.

\subsection{Temperature-driven swelling of charged microgels with explicit counterions}
\label{subsec:swelling}
In this section, we analyze in detail the deswelling behavior of the microgels with explicit counterions by adding the solvophobic potential $V_{\alpha}$ between monomers (Eq.~\ref{eq:valpha}) to mimic the increase of temperature in experiments\cite{fnieves_multiresponsive,seiffert}.

\begin{figure}[!ht]
	\includegraphics[width=0.6\textwidth]{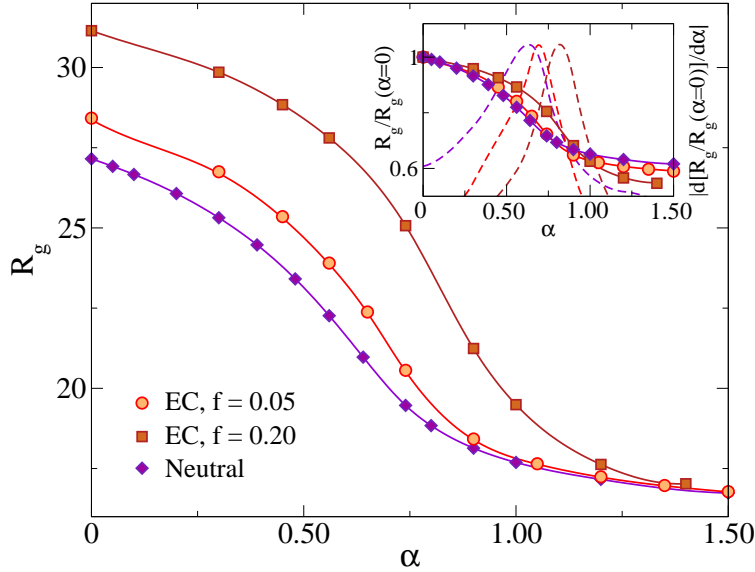}
	\caption{\label{fig:swelling2} \small Swelling curves (radius of gyration $R_g$ versus effective temperature $\alpha$) for the microgels with explicit counterions (EC) with $f=0.05$ and $f=0.20$, as compared to the neutral microgel. Inset: data normalized to the value of $R_g(\alpha=0)$ in good solvent conditions (left axis). Dashed lines report $|d[R_g/R_g(\alpha=0)]/d\alpha|$ (right axis), whose maximum corresponds to the VPT transition. These curves are arbitrarily shifted along the $y$-axis to improve visualization.}
\end{figure}

\subsubsection{Swelling curves and distribution of counterions}
We start by showing the swelling curves of the microgels in Fig.~\ref{fig:swelling2}, reporting the radius of gyration $R_g$ as a function of the parameter $\alpha$ in the presence of explicit counterions for two different values of $f$. The behavior of the neutral microgel model is also reported for comparison. The first important observation is that the value of $\alpha$ at which the VPT occurs, \textit{i.e.} $\alpha_{VPT}$, defined as the position of the maximum of $|d R_g/d\alpha|$, shifts from $\alpha_{VPT}\sim 0.63$ found in neutral microgels \cite{Moreno,andrea2019preprint}, to $\alpha_{VPT}\sim 0.69$ for $f=0.05$ and up to $\alpha_{VPT}\sim 0.82$ for $f=0.20$, as reported in the inset of Fig.~\ref{fig:swelling2}.
Using the $\alpha-T$ mapping validated against experiments for neutral PNIPAM microgels with $c=5\%$ crosslinkers and hydrodynamic radius of $\approx 400$nm\cite{andrea2019preprint}, the shifts would correspond to an increase from $\sim32^{\circ}$C for neutral microgels to $\approx 34^{\circ}$C for microgels with $f=0.05$ and $\approx 37.5^{\circ}$C for $f=0.20$, respectively. These specific values should be taken with care, since the $\alpha-T$ mapping has been validated for non-charged microgels only and may not hold in the ionic case. Regardless, the observed trend of the increase of $T_{VPT}$ with increasing charge is in qualitative agreement with experiments\cite{fnieves_multiresponsive,li2007volume}.

Additionally, we find that, while the microgel radius of gyration becomes larger with increasing charge for small values of $\alpha$, for $\alpha \gtrsim \alpha_{VPT}$ all microgels have the same size, indicating that the collapsed state does not depend on the presence of charges. This result has the interesting consequence that, upon rescaling $R_g$ by its value at the maximally swollen state ($\alpha=0$), as shown in the inset of Fig.~\ref{fig:swelling2}, the swelling ratio becomes larger as $f$ increases. Since such a ratio has been previously adopted as a measure of the particle softness\cite{mattsson,valentina}, this suggests that more charged microgels are softer than less charged or neutral ones, in agreement with experimental findings.

\begin{figure}[!ht]
	\includegraphics[width=0.6\textwidth]{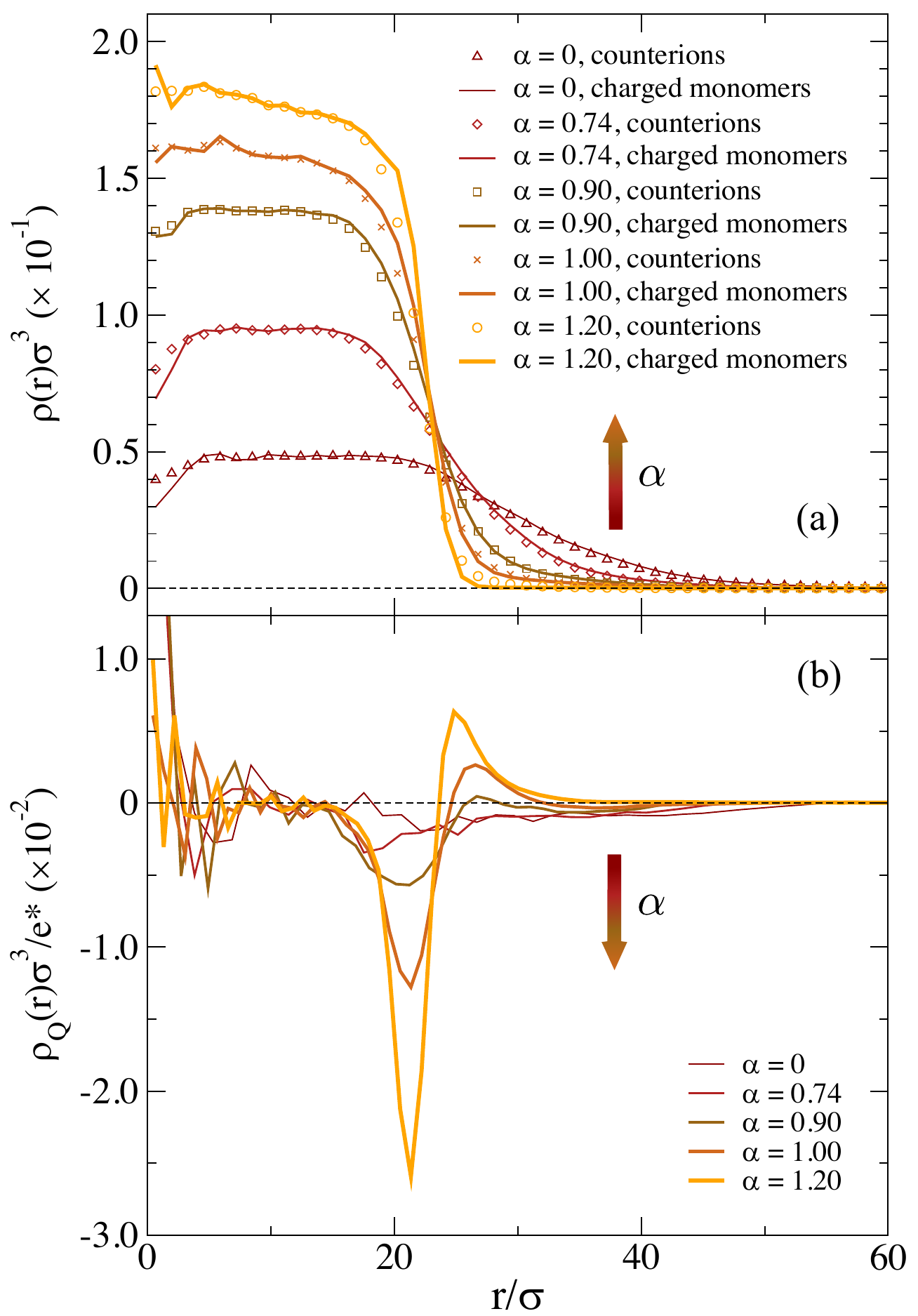}
	\caption{\label{fig:charge_alpha} \small (a) Density profiles for charged monomers (lines) and counterions (symbols) as a function of the distance from the center of mass of the microgel for $f=0.20$ from the swollen ($\alpha=0$) to the collapsed ($\alpha=1.20$) states. Included are values just below ($\alpha=0.74$) and just above ($\alpha=0.90$) the VPT; (b) net charge density profile for the same values of $\alpha$.
}
\end{figure}

The fact that the collapsed state is the same in all investigated cases could be misleadingly taken as an indication that all (or most of the) counterions are expelled from the interior of the microgel. However, this turns out not to be the case, as it can be seen from the internal charge distributions of the microgel reported in Fig.~\ref{fig:charge_alpha}. Specifically, the evolution of the charged monomers and counterions density profiles is separately shown for a few selected $\alpha$ values across the VPT in Fig.~\ref{fig:charge_alpha}(a), which only contains results for $f=0.20$. The behavior for $f=0.05$ is qualitatively similar and not shown. We find that the profiles of the charged monomers and counterions closely follow each other at all studied values of $\alpha$. This indicates a residual presence of counterions inside the microgels, which actually increases with $\alpha$
in order to balance the increase of monomer charge density in the collapsed core. The fact that the presence of counterions inside the microgels does not affect the size of the collapsed state also indirectly confirms that the choice of a small size for the counterions in our simulations is appropriate.

Looking at the profiles in Fig.~\ref{fig:charge_alpha}(a) a bit more closely, we find a small difference between counterions and charged monomers profiles upon increasing $\alpha$ and close to the surface of the microgels.  This can be better visualized in Fig.~\ref{fig:charge_alpha}(b), which reports the net-charge density profiles $\rho_Q(r)$ (defined in Methods) at different $\alpha$-values. We find that in the swollen state the charge density is statistically zero at all distances, except for the outer corona region, where it takes a tiny negative contribution. This is caused by the outermost counterions that are entropically driven to freely move around the simulation box, even far from the microgel.
This situation persists below the VPT. However a significant change occurs close and above the VPT temperature. Indeed, under these conditions the microgel still maintains a rather neutral core, but in the corona the charge density abruptly increases, leading to the formation of a charged double layer. This trend is enhanced as $\alpha$ increases, signalling that there is a large charge imbalance at the surface of the microgels, where the counterions tend to accumulate.
Such a phenomenon can be tentatively explained as follows. For small $\alpha$, the structure of the microgel is swollen and counterions can be close to the charged monomers at any distance from the center of mass, still retaining a large freedom of moving inside the network.  However, when $\alpha$ increases, the asymmetry among the interactions experienced by charged monomers and counterions come into play. On one hand, charged monomers interact with the additional solvophobic potential $V_\alpha$ which partially counteracts their electrostatic mutual repulsion. These contributions combined with the presence of the polymer network,  which constraints their positions, induce the charged monomers density $\rho_{CH}$ to steeply decay close to the surface of the microgel. On the other hand, counterions
are able to gain translational entropy and at the same time to reduce their mutual repulsion by positioning themselves close to the surface in a more dispersed way. This implies a smoother decay of $\rho_{CI}$. The asymmetry between these two behaviors causes the formation of the above-mentioned double layer. 
Hence for $\alpha > \alpha_{VPT}$, the microgels acquire an effective charge and are surrounded by a small counterion cloud. This result highlights the non-trivial arrangements of counterions with respect to the microgel structure. It can also be of guidance in the treatment of ionic microgels at finite concentrations, particularly for large ones where similar crowding effects may take place, which may cause the re-organization of the counterions distribution within the microgel even under swollen conditions\cite{ionic_schurt,denton_swelling_suspension}.

\subsection{Comparison between the explicit and implicit models across the volume phase transition}
\subsubsection{Swelling curves and snapshots}
In this section we compare the behavior of the microgels with explicit counterions with those modelled with a \dht approach across the volume phase transition for $f=0.20$. In order to make a meaningful comparison, data have been averaged over the same topologies and distributions of charged monomers.
 \begin{figure}[h]
	\includegraphics[width=0.6\textwidth]{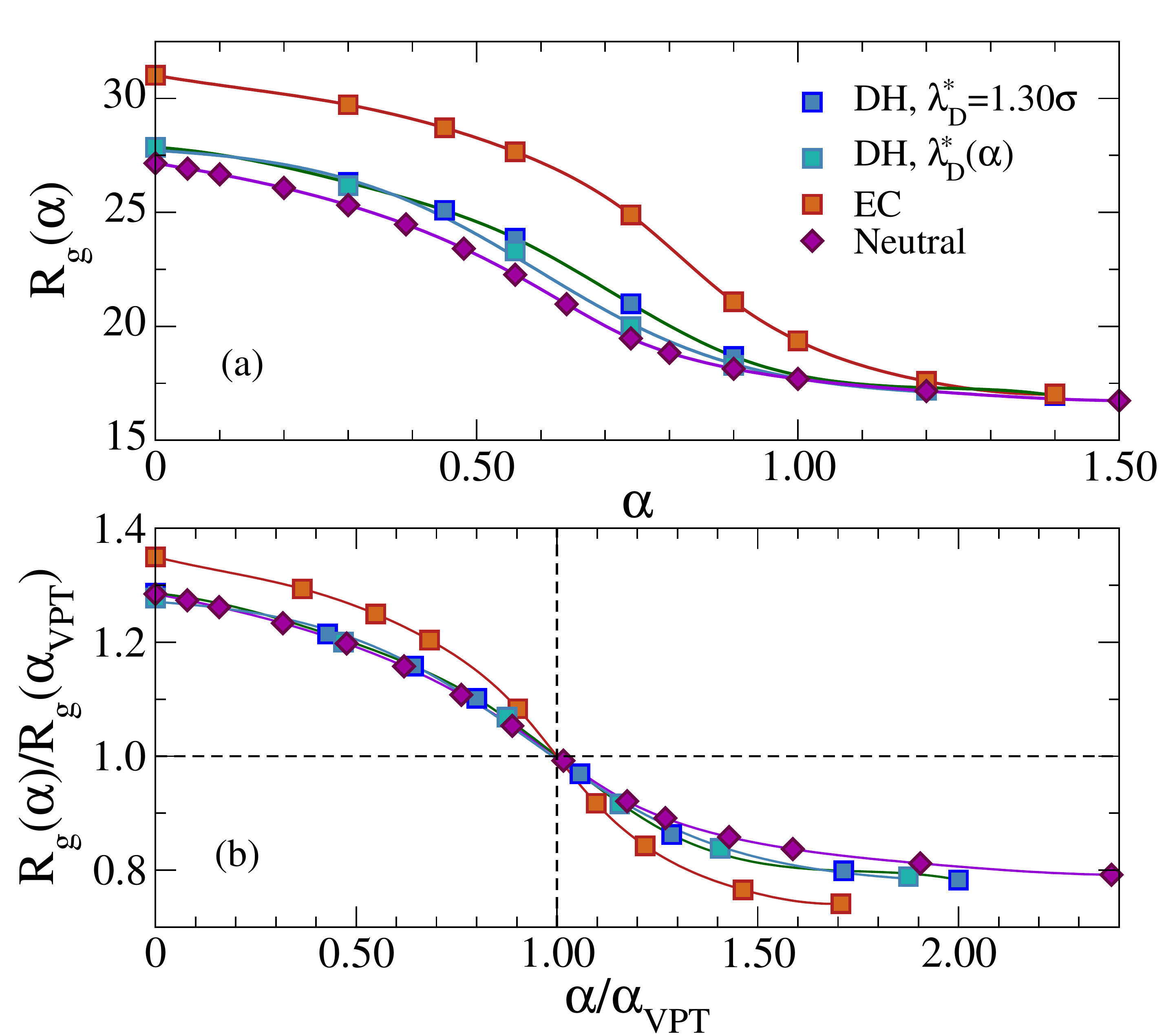}
	\caption{\label{fig:chrnd_swelling} \small (a) Swelling curves of microgels with implicit (DH) and explicit (EC) counterions for $f=0.20$, as well as the corresponding one for neutral microgels. For DH, we report both results for $\lambda^*_D=1.30\sigma$, that is the effective screening length calculated from the $\alpha=0$ microgel with explicit counterions, and for varying $\lambda^*_D(\alpha)$, calculated for each value of  $\alpha$; (b) same as in (a) but with curves rescaled by the respective values of $R_g$ and $\alpha_{VPT}$.}	
\end{figure}
We start by reporting the swelling curves of implicit and explicit models in Fig.~\ref{fig:chrnd_swelling}. For \dht simulations we have computed two different swelling curves. The first one is obtained using the effective Debye length $\ld^*$, assuming it to be constant for all values of $\alpha$. For the second swelling curve we have calculated the effective Debye length for each value of $\alpha$ to take into account the change in counterions density. 
Since the latter increases as a consequence of shrinking, the resulting $\ld^*(\alpha)$ decreases upon increasing $\alpha$.
The two swelling curves are very similar to each other, with small differences only visible close to the VPT, indicating that the transition occurs slightly earlier for the $\ld^*(\alpha)$ case with respect to the constant one. However, in both cases, $\alpha_{VPT}$ is found to be close to the neutral microgel result, and hence smaller than that of the explicit case. Interestingly, the $\ld^*(\alpha)$ case leads to $R_g$ predictions that are even further away from the explicit counterions case than those observed with a constant $\ld$, suggesting that such an approach is deeply flawed. Our findings demonstrate that the charged microgel  with explicit counterions retains a much larger structure for all $\alpha \lesssim \alpha_{VPT}$.

The bottom panel of Fig.~\ref{fig:chrnd_swelling} shows the same swelling curves rescaled along both axis with the respective values of $\alpha$ and $R_g$ at the VPT, in order to analyze the shape of the swelling curve with respect to each other. We find that, for $\alpha < \alpha_{VPT}$, both curves relative to the implicit model coincide with that of neutral microgels, while the explicit model significantly differs.  For $\alpha > \alpha_{VPT}$, on the contrary, neutral, EC and DH curves are all different and, even for the implicit model, we find that the shrinking ratio is slightly increased with respect to the neutral case, confirming that charged microgels are softer also when modelled with the \dht approach.

\begin{figure*}[h!]
  \includegraphics[width=1.0\textwidth]{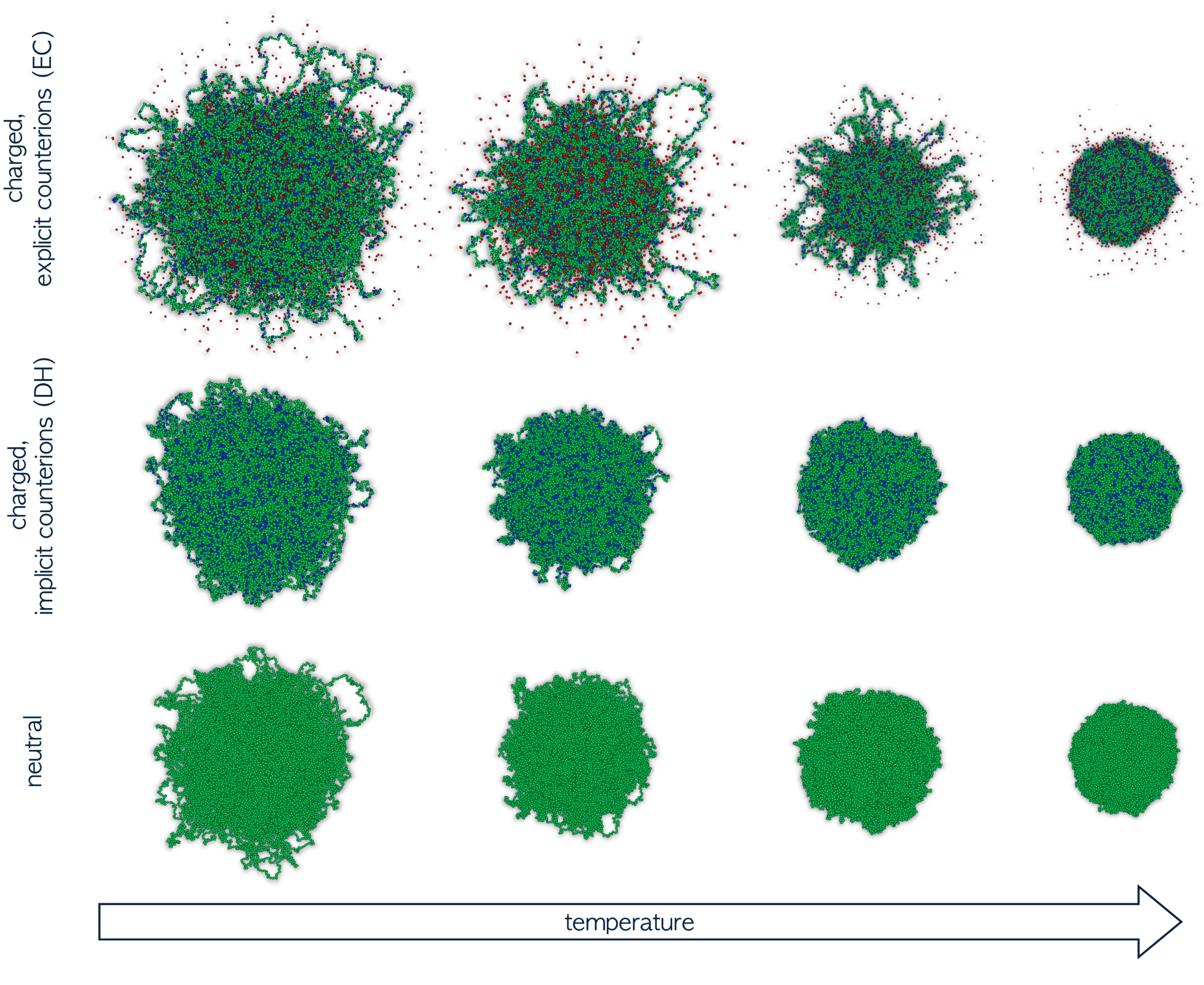}
  \caption{\label{fig:SNAPSHOTS} \small Simulation snapshots for the same microgel topology with $f=0.020$ for different models and effective temperatures. The top row shows the charged microgel with explicit counterions (EC) from left to right: in the swollen state ($\alpha=0$), just before and after the VPT ($\alpha=0.74$ and $\alpha=0.90$), and in the collapsed state ($\alpha=1.20$). The intermediate and bottom rows display corresponding states for the implicit model (DH) with $\ld^*=1.30\sigma$ ($\alpha=0.30, 0.65, 0.74$ and $1.20$) and for the neutral microgel ($\alpha=0, 0.56, 0.64$ and $1.00$), respectively. Green particles represent neutral beads, whereas the blue ones charged beads. Explicit counterions are shown as smaller red spheres.  All snapshots refer to equilibrium states, where the microgel radius of gyration fluctuates around a constant value.}
\end{figure*}

In order to better understand the main differences between the different models as the solvophobicity increases, in Fig.~\ref{fig:SNAPSHOTS} we report representative snapshots of the system across the VPT. Data for the microgel with explicit counterions (top row) are compared to the \dht model (intermediate row) and to the neutral system (bottom row) at similar values of $\alpha/\alpha_{VPT}$. All snapshots refer to the very same underlying network topology, in order to clearly discriminate the effects of charges. 
In the swollen regime, the microgel conformations are comparable, but the increase in microgel size as we go from neutral to DH to EC model is evident. By contrast, in the fully collapsed regime all microgels look very similar to each other. The most dramatic differences between the three situations can be immediately visualized close to the VPT. Under these conditions, corresponding to the second and third columns of Fig.~\ref{fig:SNAPSHOTS}, in the presence of explicit counterions the microgel appears to be made of a core and of a rather inhomogeneous corona. In fact, the most external chains do not completely collapse even when $\alpha = \alpha_{VPT}$, as they form small clusters between themselves while, at the same time, remaining clearly distinct from the homogeneous dense core. It is only when $\alpha$ significantly exceeds $\alpha_{VPT}$ that they get slowly incorporated within the core.
This behavior is completely absent both for neutral microgels and for microgels with implicit charges, where the collapse of the microgel is clearly homogeneous across the VPT, independently of the value of $\ld$. These findings can be explained by the fact that, for implicit charges, the competition between the electrostatic repulsion and the solvophobic attraction just shifts the occurrence of the VPT to larger values of $\alpha$, because a larger amount of attraction is needed to compensate the additional monomer-monomer repulsion. However, when counterions are explicitly included, they provide the system with additional degrees of freedom, thus being able to compensate the balance between attraction and repulsion even locally. This creates inhomogeneities in the charge distributions which significantly alter the microgels internal profiles, giving rise to a distinct core-corona pattern close to the VPT.

\subsubsection{Form factors}	
\begin{figure}[t]
	\includegraphics[width=0.6\textwidth]{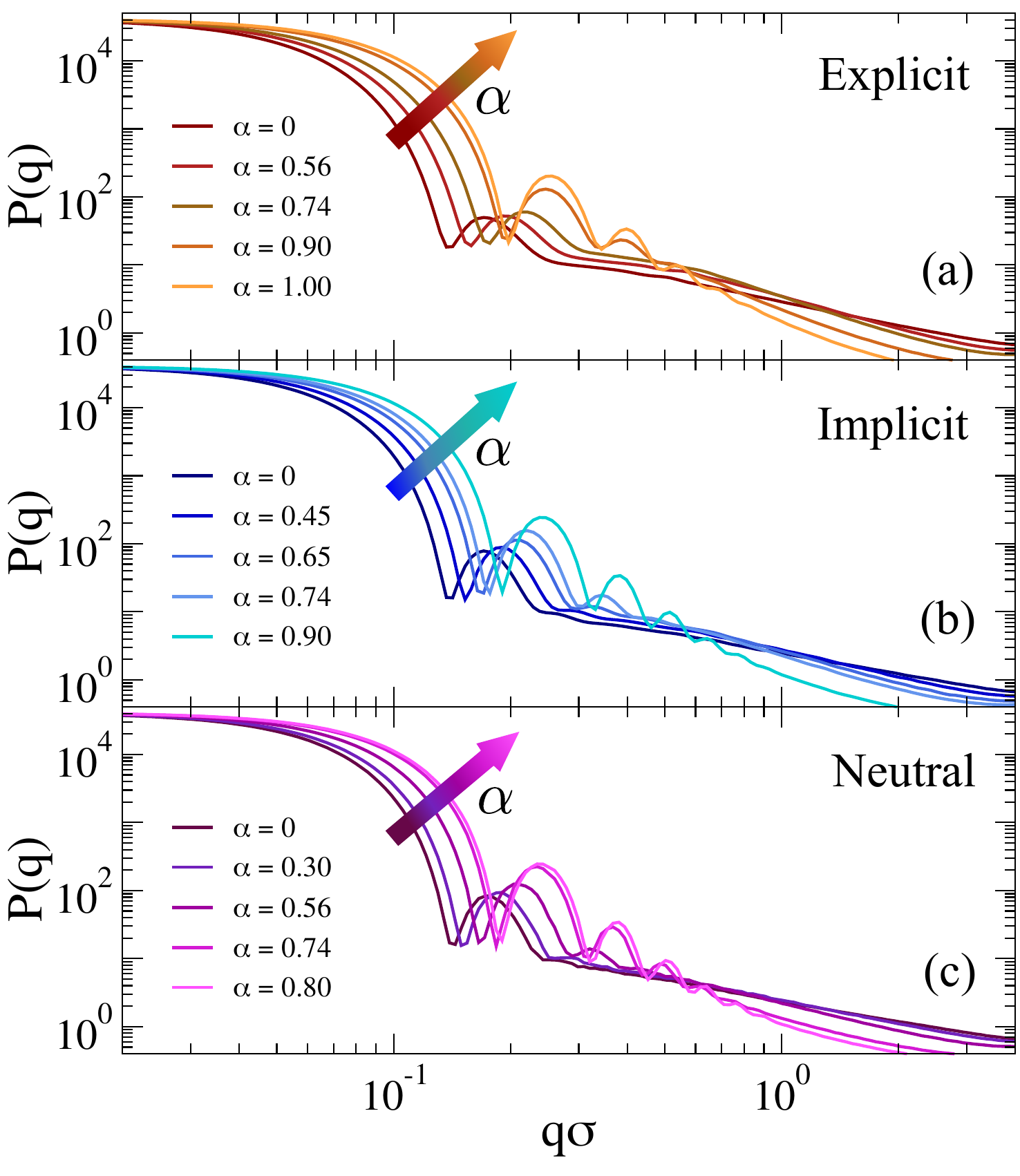}
	\caption{\label{fig:ffactors_alpha} \small Form factors for (a) explicit counterions (EC), (b) \dht (DH, $\ld=1.30\sigma$) and (c) neutral microgels with $f = 0.20$ across the VPT.}
\end{figure}

In order to better quantify the behavior observed in the snapshots, we report the form factors of the microgels in Fig.~\ref{fig:ffactors_alpha}, again comparing explicit, implicit and neutral cases at different values of $\alpha$ across the VPT.
We find evidence that the neutral and implicit cases are quite similar to each other, and both are compatible with the extended fuzzy sphere model, as shown in the Supporting Information, Fig.~S3. Instead, microgels with explicit counterions display a very different behavior in many aspects. First of all, we find that the first peak of $P(q)$ is much smaller in intensity than for the other two cases for the investigated values of $\alpha< \alpha_{VPT}$. Indeed, it tends to only shift in position without growing much in amplitude upon increasing $\alpha$. However, focusing on intermediate $q$-values beyond the first peak, $P(q)$ considerably increases in height, a feature that is absent for implicit and neutral microgels and that cannot be captured by a fuzzy-sphere-like model (see below). No secondary peaks are observed. In addition, the behavior of $P(q)$ looks almost discontinuous at the VPT temperature, sharply increasing for $\alpha > \alpha_{VPT}$ and, at the same time, developing additional peaks. As the microgel approaches the fully collapsed state, it becomes again possible to describe its form factor with the extended fuzzy sphere model.

\begin{figure}[htb]
	\includegraphics[width=0.6\textwidth]{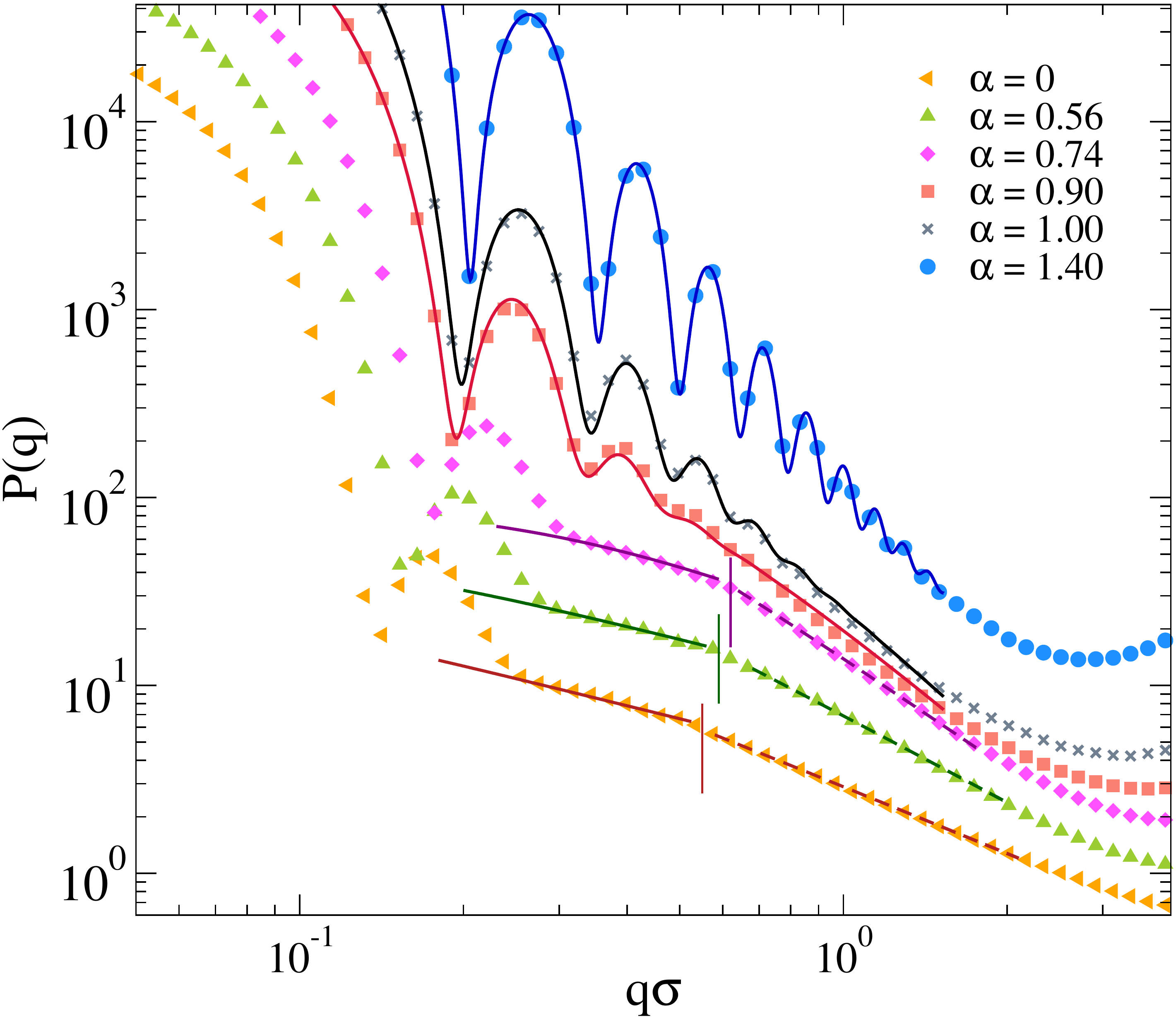}
	\caption{\label{fig:formfactors-fit} \small Zoom of the form factors of Fig.~\ref{fig:ffactors_alpha}a for microgels with explicit counterions (EC) for $f=0.20$ and relative fits. Symbols are simulation data below ($\alpha=0, 0.54, 0.76$) and above ($\alpha=0.90, 1.00, 1.40$) the VPT. Below the VPT: full lines are fits with a modified Lorentzian (Eq.\protect\ref{eq:Lorentzian}) for the $q$-range that goes from the first peak of $P(q)$ up to the correspondingly coloured vertical line;  dashed lines are fits with a different modified Lorentzian function in the range starting from the vertical line to  $q\sigma \lesssim 2$. For larger values of $q$, data are affected from finite-size effects. Above the VPT: lines are fits according to an extended fuzzy sphere model (Eq.~\protect\ref{eq:mod_fuzzy}) plus a modified Lorentzian (Eq.\protect\ref{eq:Lorentzian}).
Data sets for different $\alpha$ are arbitrarily shifted on the y-axis to improve readability.}
\end{figure}

To better discuss the features of the form factors with explicit counterions, a zoom of the data is reported in Fig.~\ref{fig:formfactors-fit}. 
For  $\alpha< \alpha_{VPT}$, where we cannot rely on a fuzzy-sphere-like model, $P(q)$ displays two distinct behaviors after the first peak, both of which are compatible with power law dependences. The first regime occurs for  $0.2 \lesssim q\sigma \lesssim 0.6$, where $P(q)\sim q^{-\delta_1}$ with the exponent $\delta_1$ being rather constant  for $\alpha< \alpha_{VPT}$, i.e. $\delta_1 = 0.75\pm0.05$. These $q$-values correspond to length scales within the corona region of the microgel. At larger $q$ the form factors exhibit a crossover to a second regime characterized by a different apparent power law.
The position of the crossover, marked with vertical lines in Fig.~\ref{fig:formfactors-fit}, shifts from $q\sigma \sim 0.55$ at $\alpha=0$ to $q\sigma \sim 0.65$ at $\alpha=0.74$.  For such second regime, a power law description of the data as $P(q)\sim q^{-\delta_2}$  gives an exponent $\delta_2$ strongly dependent on $\alpha$ (from $\sim 1.2$ at $\alpha=0$ up to $\sim 1.8$ close to the VPT). The fact that a similar power-law dependence in the first $q$-regime seems to hold for swollen microgels up to the VPT suggests that the outer corona structure remains roughly constant for this range of temperatures.  By contrast, the increase of the apparent exponent at larger $q$-values suggests that for smaller length scales the structure feels the effect of the underlying interactions, which modify the fractal properties of the network. However, at such large values of $q$, beyond $2 \sigma^{-1}$, the data suffer from finite-size effects (as it can be observed by the onset of a minimum, which precedes the occurrence of the model-dependent monomer-monomer peak at $q\sigma \sim 2\pi$\citep{camerin2018modelling}). We have thus limited our analysis here and in the following to the range $q\sigma \lesssim 2$, in which we have attempted a few types of different fits, going beyond the power-law behavior which cannot be considered to be very reliable in such a limited range of $q$ (changing by only a decade).

Among the available models, we found that the modified Lorentzian defined in Eq.~\ref{eq:Lorentzian} is able to separately describe both regimes for $\alpha< \alpha_{VPT}$, as shown in Fig.~\ref{fig:formfactors-fit}. Interestingly, the fractal exponents $D_1$ and $D_2$ extracted from the fits in the two regimes, reported in the Supporting Information (see Section ~S3), closely match the apparent power-law exponents described above. Thus, a roughly $\alpha$-independent value of $D_1$ is found for small $q$, while a larger value of $D_2$ is obtained, which rapidly increases with $\alpha$. These two parameters refer to the fractal dimensions of the correlated domains in the network over the corresponding ranges of length scales. They are coupled to two characteristic lengths, $\xi_1$ and $\xi_2$,  which quantify the correlation lengths among such domains\cite{tanaka_ffactors}. These lengths are both found to decrease with $\alpha$, in agreement with expectations. Most importantly, we find in all studied cases that $\xi_1 > \xi_2$. This suggests that the behavior of the form factors in the swollen state and up to VPT is compatible with a network with different characteristic domains  occurring in the corona and in the core region, respectively. Within the corona region (first regime), the correlation length is quite large, reflecting the few domains (visible in the snapshots of Fig.~\ref{fig:SNAPSHOTS}) that are quite far apart from each other. The fractal dimension of such environments is rather low and unaffected by changes in $\alpha$, reflecting the fact that the corona remains clearly distinct from the core, up to the VPT and beyond. Instead, within the core region (second regime), the domains correlation length is much smaller and it rapidly decreases with $\alpha$, while the fractal dimension is larger and increases with $\alpha$, consistent with the shrinking of the core. The trend of these parameters in the second regime is consistent with that found for $\alpha> \alpha_{VPT}$, in which we use an extended fuzzy sphere model plus a modified Lorentzian. Here the core-corona distinction gets less and less pronounced, suggesting that we do not need two Lorenztian terms any longer to fit the data. Further discussion on the reliability of the fits and a comparison of the extracted fit parameters with the implicit and neutral microgel models is reported in the Supporting Information.

It is interesting to compare our findings to the few available small-angle neutron scattering measurements that we are aware of, namely results for large PNIPAM-co-PAAc microgels\cite{seiffert} and for IPN microgels\cite{valentina_JCP}. In both cases, 
 the probed range of wavevectors is limited to a portion of the interior of the network, which lies well within the core region. Unfortunately, our simulated microgels are too small to make a direct comparison over a sizeable range of wave-vectors. However, in qualitative terms, both sets of experiments report an increase of the signal with $T$ for smaller values of $q$, while at larger values of the wavevectors a crossing of the data is found, so that the signal actually decreases in $T$. This seems to agree well with our numerical findings. Moreover, experimentally no peaks are detected in the probed $q$-range, and the data seems to be compatible with a power-law behavior. It remains a challenge for the future to compare our numerical form factors to light or x-ray scattering measurements that would be able to probe the corona region to confirm the lack of peaks and the power-law-like behaviors that we observe.

From all the evidence gathered in this part, we can conclude that below the VPT the competition between the solvophobic attraction and the repulsive electrostatic interactions, screened by the counterions, gives rise to a rather inhomogeneous structure, characterized by two different regimes. This complicated behaviour cannot be interpreted with a fuzzy-sphere-like description, and a new type of model would be needed to describe form factors in the whole $q$-range, perhaps inspired by multi-shell models\cite{maxime_mie,mie_lab}. On the contrary, neutral microgels and those with implicit charges display a simpler behaviour, and a modified fuzzy-sphere model with a fractal Lorentzian is sufficient to describe the data.

\subsubsection{Comparison at the same microgel size}	
\begin{figure}[t]
\includegraphics[width=1.0\textwidth]{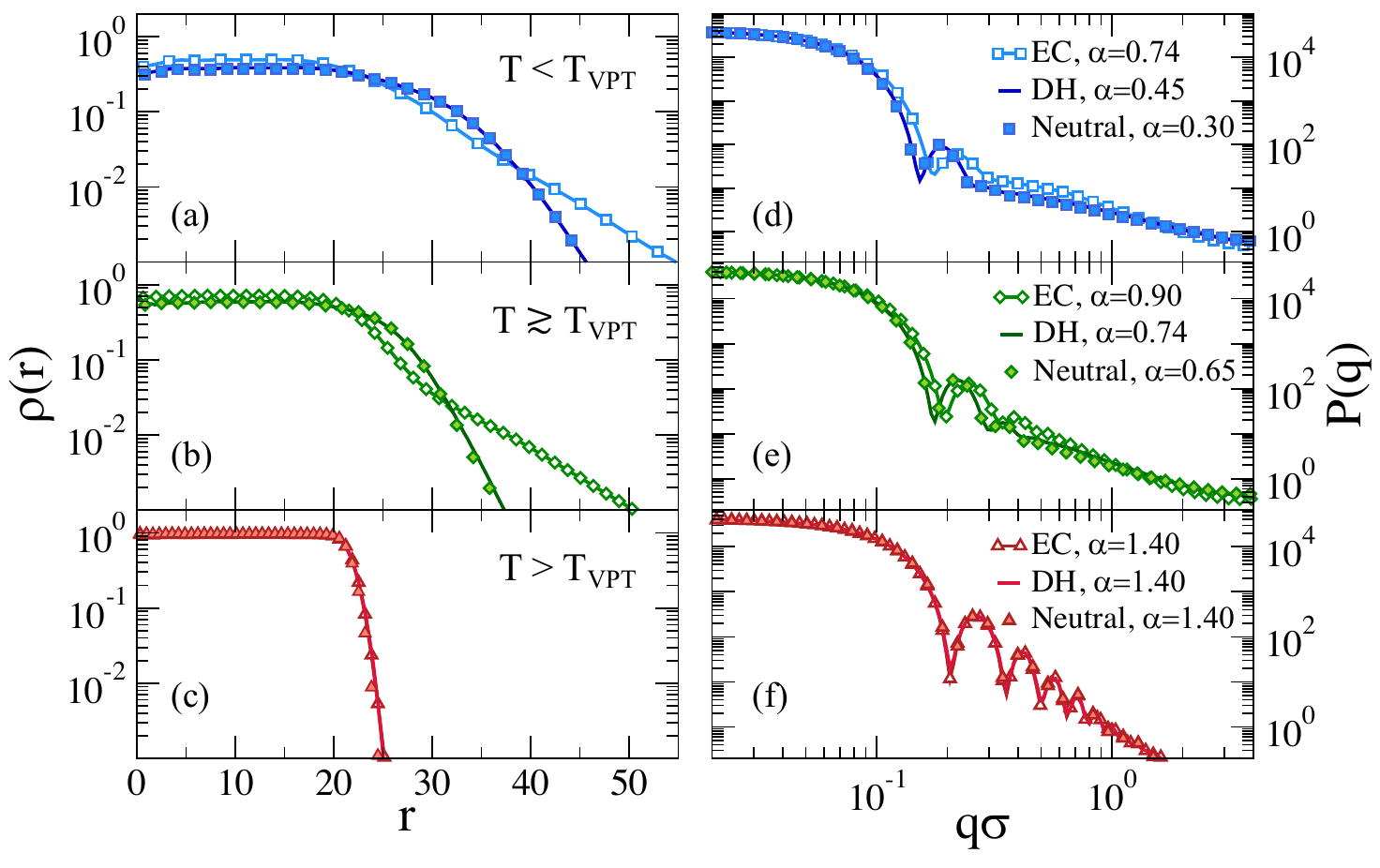}
\caption{
	\label{fig:alpha_profiles} \small (a-c) Monomer density profiles and (d-f) form factors for microgels with the same radius of gyration below ($R_g \sim 26\sigma$) and slightly above ($R_g \sim 21\sigma$) the VPT, and in the highly collapsed state ($R_g \sim 17\sigma$). The corresponding values of $\alpha$ for the three cases are: (i) $\alpha=0.74 , 0.90$ and $1.40$ for the explicit counterions (EC) model with $f=0.20$ (symbols and lines), (ii) $\alpha=0.45 , 0.74$ and $1.40$ for the \dht model (DH) with $\ld=\ld^*$ and $f=0.20$ (lines), (iii) $\alpha=0.30 , 0.65$ and $1.40$ for neutral microgels (full symbols).}
	\end{figure}

Finally, we perform a comparison for microgels with the same $R_g$ obtained with the three employed models (neutral, implicit and explicit) in order to compare differences arising in the structures when they are of roughly the same size.
We thus select the values $R_g \sim 26\sigma$, $R_g \sim 21\sigma$ and $R_g \sim 17\sigma$ for which the system is respectively below the VPT temperature, slightly above it and in the fully collapsed state. The monomer density profiles of the microgels under these conditions are reported in Fig.~\ref{fig:alpha_profiles}.

For small values of $\alpha$, we find that the monomer density inside the core is larger in the explicit charged microgels  than in neutral or implicit ones (Fig.~\ref{fig:alpha_profiles}(a,b)). This is different for what observed for the maximally swollen case ($\alpha=0$), shown in Fig.~\ref{fig:ld}(a), where the monomer density in the core was much smaller for the microgels with explicit counterions. This was due to the fact that at $\alpha=0$, the size of the microgels was very different. When we compare the models at the same $R_g$ instead, we see that the explicit counterion microgel has more monomers in the core and for large distances, while it is less dense in an intermediate range of distances. These features are maintained even when we cross the VPT, where the corona of the explicit case is much more extended than the neutral and implicit ones. Finally when the complete collapse is achieved, all the microgels have an identical density profile within numerical uncertainty (Fig.~\ref{fig:alpha_profiles}(c)). We confirm no differences at all occurring between neutral and \dht microgels at comparable $R_g$ in the investigated range of temperatures and of electrostatic parameters.

Similar plots for the form factors are reported in Fig.~\ref{fig:alpha_profiles}(d-f). Interestingly, despite the microgels having the same $R_g$, the explicit charged ones clearly show that the first peak of $P(q)$ is shifted to larger $q$-values.
This is because $R_g$ tells us how broad the mass distribution is, and we see from the density profiles that the microgels with explicit counterions have a more extended corona. This is then compensated by a smaller and denser core to produce the same $R_g$ of the other two models. We can infer that the first peak of the form factor is mainly affected by the extension of the core rather than by $R_g$. Indeed, comparing the three kinds of microgels at the same value of $\alpha$, $R_g$ is much greater for the explicit model, which thus needs a much higher intensity of the attractive force to reduce its volume. Because of the underlying inhomogeneities, this leads to a denser core, within which the screening is stronger and the attraction larger. In addition, we also confirm that the intermediate $q$ behavior of $P(q)$ is completely different for the microgel with explicit counterions, even slightly above the VPT. It is only for very large values of $\alpha$  ($\gtrsim 1.20)$ that the structure is the same for all types of microgels.

The present results further suggest that the \dht model is always very different from the one with explicit charges and that there seems to be no way to reconcile the two approaches. Thus, one could ask whether there is some other way to modify the parameters of the \dht model in order to resemble the features observed in the presence of counterions. To this aim, we would need to bypass the standard definition of the Debye length in Eq.~\ref{eq:debye}, which clearly overestimates the screening effects of the counterions present in the core. One possibility would be a phenomenological-like approach in which we consider the value of $\ld$ as the one yielding the same $R_g$ of the microgel in the maximally swollen conditions ($\alpha=0$). From Fig.~\ref{fig:chrnd_RgvsLd}, we observe that this would be achieved with a much larger value of the Debye length with respect to the effective one, i.e. $\ld=3.0\sigma$. We have thus performed additional simulations (not shown) of the \dht model for this value of $\ld$ as a function of $\alpha$ to try to assess whether in this case the implicit treatment of the charges can give rise to inhomogeneous effects such as with explicit counterions. However, we find that the microgel undergoes a microphase separation at large $\alpha$ and does not resemble at all the case with explicit counterions. We will thus address the case of microgels with very high charge fractions in future works, concluding for the present study that the \dht model yields results that appear to be too similar to those of neutral microgels, pointing to the crucial role of counterions in a correct treatment of single-microgel properties.

\section{Conclusions and perspectives}
\label{sec:conclusions}
In this work, we carried out extensive numerical investigations of charged microgels, focusing on the single-particle structure and swelling behavior across the volume phase transition. Extending a realistic assembly protocol that we recently put forward in Refs.\cite{silicomicrogel,andrea2019preprint}, we now additionally included the effects of charges in two different ways. On one hand, we employed an implicit model where screening effects of the counterions are included in a \dht treatment where we varied both the amount of charged monomers and the Debye length. On the other hand, we performed simulations in the presence of explicit counterions, interacting with the charged monomers through a Coulomb repulsion.  In both frameworks, we addressed the importance of having an underlying disordered network topology with the desired core-corona architecture, similar to that featured in microgels synthesised through the most common routes.

Our results are consistent with common expectations for the behavior of thermoresponsive microgels where charged co-monomers are included in the synthesis. In particular, we find that the size of the microgels in the swollen state increases with the fraction of charged monomers included in the network.  Such an increase is also responsible for the occurrence of larger swelling ratios for more charged microgels, confirming the link between charge and softness\cite{mattsson}. This is found for both explicit and implicit counterion modeling, thus being a robust feature of charged microgels. We also confirm that the VPT temperature shifts to larger values as the amount of charge increases, in agreement with experimental results\cite{fnieves_multiresponsive,seiffert,li2007volume}. However, some of our findings are less obvious than could be naively thought. First of all, while charged microgels are very different with respect to the neutral case below and at the VPT temperature, we find no difference among their collapsed structures, independently on the presence of charges and of the treatment of the counterions, suggesting that at sufficiently high temperatures they eventually reach a homogeneous spherical structure of the same size. Furthermore, specific considerations have been made possible by the use of explicit counterions. In particular, we find that the fully collapsed microgel in the explicit model is not at all free of counterions, which therefore are not expelled from the interior of the microgel upon deswelling. Instead, they are retained inside it
in order to balance the increased charge density of the collapsed structure. Thus, counterions freely permeate and screen monomer charges at all temperatures, acting as neutralizers for the polymer network.
Only close to the microgel surface we observe the onset of a non-neutral local charge, which manifests just above the VPT temperature in agreement with electrophoretic measurements\cite{truzzolillo_sennato}. As a consequence, charged microgels behave as overall neutral objects up to around the VPT temperature, at least in the dilute regime.

We also compared the structure and the swelling of the microgel using the \dht model, this being a much more convenient way to treat charges from the theoretical and numerical point of view. It turned out that a qualitative agreement between the explicit and the implicit approaches is unachievable, even using an effective Debye length that was calculated from the density of counterions obtained within the explicit case. Our findings indicate that the \dht approach is not able to reproduce many important effects that arise in the presence of charges, being mainly able to describe the average effect of screening of counterions onto charged beads over the polymer network. In particular, it fails to take into account the osmotic pressure of both inner counterions, acting in favour of the microgel swelling, and external ones, acting against the swelling. 
Remarkably, the structural features observed for the \dht model are actually much more similar to those of neutral microgels than to the explicit counterions case. The most prominent difference can be noticed in the snapshots of Fig.~\ref{fig:SNAPSHOTS}, where the inhomogeneous core-corona structure is augmented by the presence of charges and counterions, a fact that is completely missing in the implicit representation. Strikingly, this reflects on the form factors of the microgels, which show a profile that is incompatible with the a fuzzy-sphere-like model, but rather display the onset of two distinct regimes, each of them compatible with a modified Lorentzian. Gaining a strong theoretical understanding of these intriguing findings will be the subject of future work. A potentially interesting perspective would be to combine our simulations with theoretical approaches\cite{moncho_OZ,moncho_DFT} in order to provide some description of the data and perhaps to develop a modified \dht approach, which could take into account the inhomogeneity of the microgel, assigning different values of $\ld$ to the core and to the corona, respectively. However, how to determine these values a priori (\textit{i.e.} without estimating them with explicit simulations) remains an open question.

The understanding of the single-particle properties of co-polymerised microgels can be considered as a first step toward a better understanding of interpenetrated network microgels (IPN), wherein PNIPAM and PAAc are organized into two independent,  interpenetrated networks, so that the responsiveness to temperature and to pH can be decoupled\cite{nigro2015dynamic}. This particular kind of microgels has recently gathered a lot of interest because of their intriguing fragility behavior: as the amount of charges increase, these systems exhibit features of strong glass-formers, a rather unique example in soft matter \cite{mattsson,valentina}. A recent work\cite{philippe}
 has put forward the idea that this behavior directly stems from charge effects, which also increase the softness of the particles, as confirmed in the present work. It would thus be very interesting to address the behavior of IPN microgels in future works.

Finally, our aim will be to transfer the knowledge from single-particle properties to many-body systems by developing appropriate coarse-grained effective potentials, still retaining the essential ingredients of the microgels, in order to be able to address their structural and dynamical behavior at various concentrations. Hence, by calculating the effective potential between two charged microgels we could validate and refine the effective approaches carried out in recent works on the assembly properties of charged microgels in bulk\cite{ionic_selfassembly_2}.

\section*{Acknowledgments} 
We thank J. Ruiz Franco for valuable discussions. We acknowledge support from the European Research Council (ERC Consolidator Grant 681597, MIMIC). 

\section*{Supporting Information}
Assessment of the size effects by varying the side of the simulation box, preliminary analysis on the choice of the counterion size, discussion on the form factors fits and corresponding parameters for the models analyzed in the main text.

\bibliography{bibliography}

\providecommand{\noopsort}[1]{}\providecommand{\singleletter}[1]{#1}%
\providecommand{\latin}[1]{#1}
\makeatletter
\providecommand{\doi}
  {\begingroup\let\do\@makeother\dospecials
  \catcode`\{=1 \catcode`\}=2 \doi@aux}
\providecommand{\doi@aux}[1]{\endgroup\texttt{#1}}
\makeatother
\providecommand*\mcitethebibliography{\thebibliography}
\csname @ifundefined\endcsname{endmcitethebibliography}
  {\let\endmcitethebibliography\endthebibliography}{}
\begin{mcitethebibliography}{61}
\providecommand*\natexlab[1]{#1}
\providecommand*\mciteSetBstSublistMode[1]{}
\providecommand*\mciteSetBstMaxWidthForm[2]{}
\providecommand*\mciteBstWouldAddEndPuncttrue
  {\def\EndOfBibitem{\unskip.}}
\providecommand*\mciteBstWouldAddEndPunctfalse
  {\let\EndOfBibitem\relax}
\providecommand*\mciteSetBstMidEndSepPunct[3]{}
\providecommand*\mciteSetBstSublistLabelBeginEnd[3]{}
\providecommand*\EndOfBibitem{}
\mciteSetBstSublistMode{f}
\mciteSetBstMaxWidthForm{subitem}{(\alph{mcitesubitemcount})}
\mciteSetBstSublistLabelBeginEnd
  {\mcitemaxwidthsubitemform\space}
  {\relax}
  {\relax}

\bibitem[Yunker \latin{et~al.}(2014)Yunker, Chen, Gratale, Lohr, Still, and
  Yodh]{yunker2014physics}
Yunker,~P.~J.; Chen,~K.; Gratale,~M.~D.; Lohr,~M.~A.; Still,~T.; Yodh,~A.
  Physics in ordered and disordered colloidal matter composed of poly
  (N-isopropylacrylamide) microgel particles. \emph{Reports on Progress in
  Physics} \textbf{2014}, \emph{77}, 056601\relax
\mciteBstWouldAddEndPuncttrue
\mciteSetBstMidEndSepPunct{\mcitedefaultmidpunct}
{\mcitedefaultendpunct}{\mcitedefaultseppunct}\relax
\EndOfBibitem
\bibitem[Lyon and Fernandez-Nieves(2012)Lyon, and
  Fernandez-Nieves]{fnieves_lyon}
Lyon,~L.~A.; Fernandez-Nieves,~A. The polymer/colloid duality of microgel
  suspensions. \emph{Annual review of physical chemistry} \textbf{2012},
  \emph{63}, 25--43\relax
\mciteBstWouldAddEndPuncttrue
\mciteSetBstMidEndSepPunct{\mcitedefaultmidpunct}
{\mcitedefaultendpunct}{\mcitedefaultseppunct}\relax
\EndOfBibitem
\bibitem[Brijitta and Schurtenberger(2019)Brijitta, and
  Schurtenberger]{brijitta2019responsive}
Brijitta,~J.; Schurtenberger,~P. Responsive Hydrogel Colloids: Structure,
  Interactions, Phase Behaviour, and Equilibrium and Non-Equilibrium
  Transitions of Microgel Dispersions. \emph{Current Opinion in Colloid \&
  Interface Science} \textbf{2019}, DOI: 10.1016/j.cocis.2019.02.005\relax
\mciteBstWouldAddEndPuncttrue
\mciteSetBstMidEndSepPunct{\mcitedefaultmidpunct}
{\mcitedefaultendpunct}{\mcitedefaultseppunct}\relax
\EndOfBibitem
\bibitem[Karg \latin{et~al.}(2019)Karg, Pich, Hellweg, Hoare, Lyon, Crassous,
  Suzuki, Gumerov, Schneider, Potemkin, \latin{et~al.}
  others]{karg2019nanogels}
Karg,~M.; Pich,~A.; Hellweg,~T.; Hoare,~T.; Lyon,~L.~A.; Crassous,~J.~J.;
  Suzuki,~D.; Gumerov,~R.~A.; Schneider,~S.; Potemkin,~I.~I. \latin{et~al.}
  Nanogels and microgels: From model colloids to applications, recent
  developments and future trends. \emph{Langmuir} \textbf{2019}, DOI:
  10.1021/acs.langmuir.8b04304\relax
\mciteBstWouldAddEndPuncttrue
\mciteSetBstMidEndSepPunct{\mcitedefaultmidpunct}
{\mcitedefaultendpunct}{\mcitedefaultseppunct}\relax
\EndOfBibitem
\bibitem[Fernandez-Nieves \latin{et~al.}(2011)Fernandez-Nieves, Wyss, Mattsson,
  and Weitz]{fernandez2011microgel}
Fernandez-Nieves,~A.; Wyss,~H.; Mattsson,~J.; Weitz,~D.~A. \emph{Microgel
  suspensions: fundamentals and applications}; John Wiley \& Sons, 2011\relax
\mciteBstWouldAddEndPuncttrue
\mciteSetBstMidEndSepPunct{\mcitedefaultmidpunct}
{\mcitedefaultendpunct}{\mcitedefaultseppunct}\relax
\EndOfBibitem
\bibitem[Pelton and Hoare(2011)Pelton, and Hoare]{synthesis}
Pelton,~R.; Hoare,~T. Microgels and their synthesis: An introduction.
  \emph{Microgel Suspensions: Fundamentals and Applications} \textbf{2011},
  \emph{1}, 1--32\relax
\mciteBstWouldAddEndPuncttrue
\mciteSetBstMidEndSepPunct{\mcitedefaultmidpunct}
{\mcitedefaultendpunct}{\mcitedefaultseppunct}\relax
\EndOfBibitem
\bibitem[N{\"o}jd \latin{et~al.}(2018)N{\"o}jd, Holmqvist, Boon, Obiols-Rabasa,
  Mohanty, Schweins, and Schurtenberger]{ionic_schurt}
N{\"o}jd,~S.; Holmqvist,~P.; Boon,~N.; Obiols-Rabasa,~M.; Mohanty,~P.~S.;
  Schweins,~R.; Schurtenberger,~P. Deswelling behaviour of ionic microgel
  particles from low to ultra-high densities. \emph{Soft matter} \textbf{2018},
  \emph{14}, 4150--4159\relax
\mciteBstWouldAddEndPuncttrue
\mciteSetBstMidEndSepPunct{\mcitedefaultmidpunct}
{\mcitedefaultendpunct}{\mcitedefaultseppunct}\relax
\EndOfBibitem
\bibitem[Rochette \latin{et~al.}(2017)Rochette, Kent, Habicht, and
  Seiffert]{seiffert}
Rochette,~D.; Kent,~B.; Habicht,~A.; Seiffert,~S. Effect of polymer network
  inhomogeneity on the volume phase transitions of thermo-and pH-sensitive
  weakly charged microgels. \emph{Colloid and Polymer Science} \textbf{2017},
  \emph{295}, 507--520\relax
\mciteBstWouldAddEndPuncttrue
\mciteSetBstMidEndSepPunct{\mcitedefaultmidpunct}
{\mcitedefaultendpunct}{\mcitedefaultseppunct}\relax
\EndOfBibitem
\bibitem[Capriles-Gonz{\'a}lez \latin{et~al.}(2008)Capriles-Gonz{\'a}lez,
  Sierra-Mart{\'\i}n, Fern{\'a}ndez-Nieves, and
  Fern{\'a}ndez-Barbero]{fnieves_multiresponsive}
Capriles-Gonz{\'a}lez,~D.; Sierra-Mart{\'\i}n,~B.; Fern{\'a}ndez-Nieves,~A.;
  Fern{\'a}ndez-Barbero,~A. Coupled deswelling of multiresponse microgels.
  \emph{The Journal of Physical Chemistry B} \textbf{2008}, \emph{112},
  12195--12200\relax
\mciteBstWouldAddEndPuncttrue
\mciteSetBstMidEndSepPunct{\mcitedefaultmidpunct}
{\mcitedefaultendpunct}{\mcitedefaultseppunct}\relax
\EndOfBibitem
\bibitem[Truzzolillo \latin{et~al.}(2018)Truzzolillo, Sennato, Sarti,
  Casciardi, Bazzoni, and Bordi]{truzzolillo_sennato}
Truzzolillo,~D.; Sennato,~S.; Sarti,~S.; Casciardi,~S.; Bazzoni,~C.; Bordi,~F.
  Overcharging and reentrant condensation of thermoresponsive ionic microgels.
  \emph{Soft matter} \textbf{2018}, \emph{14}, 4110--4125\relax
\mciteBstWouldAddEndPuncttrue
\mciteSetBstMidEndSepPunct{\mcitedefaultmidpunct}
{\mcitedefaultendpunct}{\mcitedefaultseppunct}\relax
\EndOfBibitem
\bibitem[N{\"o}jd \latin{et~al.}(2013)N{\"o}jd, Mohanty, Bagheri, Yethiraj, and
  Schurtenberger]{ionic_selfassembly_1}
N{\"o}jd,~S.; Mohanty,~P.~S.; Bagheri,~P.; Yethiraj,~A.; Schurtenberger,~P.
  Electric field driven self-assembly of ionic microgels. \emph{Soft Matter}
  \textbf{2013}, \emph{9}, 9199--9207\relax
\mciteBstWouldAddEndPuncttrue
\mciteSetBstMidEndSepPunct{\mcitedefaultmidpunct}
{\mcitedefaultendpunct}{\mcitedefaultseppunct}\relax
\EndOfBibitem
\bibitem[Colla \latin{et~al.}(2018)Colla, Mohanty, N{\"o}jd, Bialik, Riede,
  Schurtenberger, and Likos]{ionic_selfassembly_2}
Colla,~T.; Mohanty,~P.~S.; N{\"o}jd,~S.; Bialik,~E.; Riede,~A.;
  Schurtenberger,~P.; Likos,~C.~N. Self-Assembly of Ionic Microgels Driven by
  an Alternating Electric Field: Theory, Simulations, and Experiments.
  \emph{ACS nano} \textbf{2018}, \emph{12}, 4321--4337\relax
\mciteBstWouldAddEndPuncttrue
\mciteSetBstMidEndSepPunct{\mcitedefaultmidpunct}
{\mcitedefaultendpunct}{\mcitedefaultseppunct}\relax
\EndOfBibitem
\bibitem[Eichenbaum \latin{et~al.}(1999)Eichenbaum, Kiser, Dobrynin, Simon, and
  Needham]{uptake_release}
Eichenbaum,~G.~M.; Kiser,~P.~F.; Dobrynin,~A.~V.; Simon,~S.~A.; Needham,~D.
  Investigation of the Swelling Response and Loading of Ionic Microgels with
  Drugs and Proteins: The Dependence on Cross-Link Density.
  \emph{Macromolecules} \textbf{1999}, \emph{32}, 4867--4878\relax
\mciteBstWouldAddEndPuncttrue
\mciteSetBstMidEndSepPunct{\mcitedefaultmidpunct}
{\mcitedefaultendpunct}{\mcitedefaultseppunct}\relax
\EndOfBibitem
\bibitem[Reese \latin{et~al.}(2004)Reese, Mikhonin, Kamenjicki, Tikhonov, and
  Asher]{photonic_crystal}
Reese,~C.~E.; Mikhonin,~A.~V.; Kamenjicki,~M.; Tikhonov,~A.; Asher,~S.~A.
  Nanogel nanosecond photonic crystal optical switching. \emph{Journal of the
  American Chemical Society} \textbf{2004}, \emph{126}, 1493--1496\relax
\mciteBstWouldAddEndPuncttrue
\mciteSetBstMidEndSepPunct{\mcitedefaultmidpunct}
{\mcitedefaultendpunct}{\mcitedefaultseppunct}\relax
\EndOfBibitem
\bibitem[Howe \latin{et~al.}(2009)Howe, Desrousseaux, Lunel, Tavacoli, Yow, and
  Routh]{howe}
Howe,~A.~M.; Desrousseaux,~S.; Lunel,~L.~S.; Tavacoli,~J.; Yow,~H.~N.;
  Routh,~A.~F. Anomalous viscosity jump during the volume phase transition of
  poly (N-isopropylacrylamide) particles. \emph{Advances in colloid and
  interface science} \textbf{2009}, \emph{147}, 124--131\relax
\mciteBstWouldAddEndPuncttrue
\mciteSetBstMidEndSepPunct{\mcitedefaultmidpunct}
{\mcitedefaultendpunct}{\mcitedefaultseppunct}\relax
\EndOfBibitem
\bibitem[Mattsson \latin{et~al.}(2009)Mattsson, Wyss, Fernandez-Nieves,
  Miyazaki, Hu, Reichman, and Weitz]{mattsson}
Mattsson,~J.; Wyss,~H.~M.; Fernandez-Nieves,~A.; Miyazaki,~K.; Hu,~Z.;
  Reichman,~D.~R.; Weitz,~D.~A. Soft colloids make strong glasses.
  \emph{Nature} \textbf{2009}, \emph{462}, 83\relax
\mciteBstWouldAddEndPuncttrue
\mciteSetBstMidEndSepPunct{\mcitedefaultmidpunct}
{\mcitedefaultendpunct}{\mcitedefaultseppunct}\relax
\EndOfBibitem
\bibitem[Nigro \latin{et~al.}(2017)Nigro, Angelini, Bertoldo, Bruni, Ricci, and
  Ruzicka]{valentina}
Nigro,~V.; Angelini,~R.; Bertoldo,~M.; Bruni,~F.; Ricci,~M.~A.; Ruzicka,~B.
  Dynamical behavior of microgels of interpenetrated polymer networks.
  \emph{Soft matter} \textbf{2017}, \emph{13}, 5185--5193\relax
\mciteBstWouldAddEndPuncttrue
\mciteSetBstMidEndSepPunct{\mcitedefaultmidpunct}
{\mcitedefaultendpunct}{\mcitedefaultseppunct}\relax
\EndOfBibitem
\bibitem[De~Gennes(1979)]{degennes}
De~Gennes,~P.~G. \emph{Scaling concepts in polymer physics}; Cornell university
  press, 1979\relax
\mciteBstWouldAddEndPuncttrue
\mciteSetBstMidEndSepPunct{\mcitedefaultmidpunct}
{\mcitedefaultendpunct}{\mcitedefaultseppunct}\relax
\EndOfBibitem
\bibitem[Denton and Tang(2016)Denton, and Tang]{denton2}
Denton,~A.~R.; Tang,~Q. Counterion-induced swelling of ionic microgels.
  \emph{The Journal of chemical physics} \textbf{2016}, \emph{145},
  164901\relax
\mciteBstWouldAddEndPuncttrue
\mciteSetBstMidEndSepPunct{\mcitedefaultmidpunct}
{\mcitedefaultendpunct}{\mcitedefaultseppunct}\relax
\EndOfBibitem
\bibitem[Moncho-Jord{\'a}(2013)]{moncho_OZ}
Moncho-Jord{\'a},~A. Effective charge of ionic microgel particles in the
  swollen and collapsed states: The role of the steric microgel-ion repulsion.
  \emph{The Journal of chemical physics} \textbf{2013}, \emph{139},
  064906\relax
\mciteBstWouldAddEndPuncttrue
\mciteSetBstMidEndSepPunct{\mcitedefaultmidpunct}
{\mcitedefaultendpunct}{\mcitedefaultseppunct}\relax
\EndOfBibitem
\bibitem[Moncho-Jord{\'a} and Dzubiella(2016)Moncho-Jord{\'a}, and
  Dzubiella]{moncho_DFT}
Moncho-Jord{\'a},~A.; Dzubiella,~J. Swelling of ionic microgel particles in the
  presence of excluded-volume interactions: a density functional approach.
  \emph{Physical Chemistry Chemical Physics} \textbf{2016}, \emph{18},
  5372--5385\relax
\mciteBstWouldAddEndPuncttrue
\mciteSetBstMidEndSepPunct{\mcitedefaultmidpunct}
{\mcitedefaultendpunct}{\mcitedefaultseppunct}\relax
\EndOfBibitem
\bibitem[Denton(2003)]{denton1}
Denton,~A. Counterion penetration and effective electrostatic interactions in
  solutions of polyelectrolyte stars and microgels. \emph{Physical Review E}
  \textbf{2003}, \emph{67}, 011804\relax
\mciteBstWouldAddEndPuncttrue
\mciteSetBstMidEndSepPunct{\mcitedefaultmidpunct}
{\mcitedefaultendpunct}{\mcitedefaultseppunct}\relax
\EndOfBibitem
\bibitem[Likos(2011)]{likos_mgel_book}
Likos,~C.~N. Structure and thermodynamics of ionic microgels. \emph{Microgel
  Suspensions: Fundamentals and Applications} \textbf{2011}, 163--193\relax
\mciteBstWouldAddEndPuncttrue
\mciteSetBstMidEndSepPunct{\mcitedefaultmidpunct}
{\mcitedefaultendpunct}{\mcitedefaultseppunct}\relax
\EndOfBibitem
\bibitem[Weyer and Denton(2018)Weyer, and Denton]{denton_swelling_suspension}
Weyer,~T.~J.; Denton,~A.~R. Concentration-dependent swelling and structure of
  ionic microgels: simulation and theory of a coarse-grained model. \emph{Soft
  Matter} \textbf{2018}, \emph{14}, 4530--4540\relax
\mciteBstWouldAddEndPuncttrue
\mciteSetBstMidEndSepPunct{\mcitedefaultmidpunct}
{\mcitedefaultendpunct}{\mcitedefaultseppunct}\relax
\EndOfBibitem
\bibitem[Rovigatti \latin{et~al.}(2019)Rovigatti, Gnan, Tavagnacco, Moreno, and
  Zaccarelli]{rovigatti_review}
Rovigatti,~L.; Gnan,~N.; Tavagnacco,~L.; Moreno,~A.~J.; Zaccarelli,~E.
  Numerical modelling of non-ionic microgels: an overview. \emph{Soft matter}
  \textbf{2019}, \emph{15}, 1108--1119\relax
\mciteBstWouldAddEndPuncttrue
\mciteSetBstMidEndSepPunct{\mcitedefaultmidpunct}
{\mcitedefaultendpunct}{\mcitedefaultseppunct}\relax
\EndOfBibitem
\bibitem[Mart{\'\i}n-Molina and Quesada-P{\'e}rez(2019)Mart{\'\i}n-Molina, and
  Quesada-P{\'e}rez]{quesada_review}
Mart{\'\i}n-Molina,~A.; Quesada-P{\'e}rez,~M. A review of coarse-grained
  simulations of nanogel and microgel particles. \emph{Journal of Molecular
  Liquids} \textbf{2019}, DOI: 10.1016/j.molliq.2019.02.030\relax
\mciteBstWouldAddEndPuncttrue
\mciteSetBstMidEndSepPunct{\mcitedefaultmidpunct}
{\mcitedefaultendpunct}{\mcitedefaultseppunct}\relax
\EndOfBibitem
\bibitem[Kobayashi and Winkler(2014)Kobayashi, and Winkler]{winklerDH}
Kobayashi,~H.; Winkler,~R.~G. Structure of Microgels with Debye–Hückel
  Interactions. \emph{Polymers} \textbf{2014}, \emph{6}, 1602--1617\relax
\mciteBstWouldAddEndPuncttrue
\mciteSetBstMidEndSepPunct{\mcitedefaultmidpunct}
{\mcitedefaultendpunct}{\mcitedefaultseppunct}\relax
\EndOfBibitem
\bibitem[Kobayashi \latin{et~al.}(2017)Kobayashi, Halver, Sutmann, and
  Winkler]{kobayashi_salt}
Kobayashi,~H.; Halver,~R.; Sutmann,~G.; Winkler,~R.~G. Polymer conformations in
  ionic microgels in the presence of salt: Theoretical and mesoscale simulation
  results. \emph{Polymers} \textbf{2017}, \emph{9}, 15\relax
\mciteBstWouldAddEndPuncttrue
\mciteSetBstMidEndSepPunct{\mcitedefaultmidpunct}
{\mcitedefaultendpunct}{\mcitedefaultseppunct}\relax
\EndOfBibitem
\bibitem[Schneider and Linse(2003)Schneider, and Linse]{linse_hydrogel}
Schneider,~S.; Linse,~P. Monte Carlo simulation of defect-free cross-linked
  polyelectrolyte gels. \emph{The Journal of Physical Chemistry B}
  \textbf{2003}, \emph{107}, 8030--8040\relax
\mciteBstWouldAddEndPuncttrue
\mciteSetBstMidEndSepPunct{\mcitedefaultmidpunct}
{\mcitedefaultendpunct}{\mcitedefaultseppunct}\relax
\EndOfBibitem
\bibitem[Claudio \latin{et~al.}(2009)Claudio, Kremer, and Holm]{holm_cions}
Claudio,~G.~C.; Kremer,~K.; Holm,~C. Comparison of a hydrogel model to the
  Poisson--Boltzmann cell model. \emph{The Journal of chemical physics}
  \textbf{2009}, \emph{131}, 094903\relax
\mciteBstWouldAddEndPuncttrue
\mciteSetBstMidEndSepPunct{\mcitedefaultmidpunct}
{\mcitedefaultendpunct}{\mcitedefaultseppunct}\relax
\EndOfBibitem
\bibitem[Quesada-P{\'e}rez \latin{et~al.}(2012)Quesada-P{\'e}rez, Ramos,
  Forcada, and Mart{\'\i}n-Molina]{quesada2012cions}
Quesada-P{\'e}rez,~M.; Ramos,~J.; Forcada,~J.; Mart{\'\i}n-Molina,~A. Computer
  simulations of thermo-sensitive microgels: Quantitative comparison with
  experimental swelling data. \emph{The Journal of chemical physics}
  \textbf{2012}, \emph{136}, 244903\relax
\mciteBstWouldAddEndPuncttrue
\mciteSetBstMidEndSepPunct{\mcitedefaultmidpunct}
{\mcitedefaultendpunct}{\mcitedefaultseppunct}\relax
\EndOfBibitem
\bibitem[Quesada-P{\'e}rez and Mart{\'\i}n-Molina(2013)Quesada-P{\'e}rez, and
  Mart{\'\i}n-Molina]{quesada2013cions}
Quesada-P{\'e}rez,~M.; Mart{\'\i}n-Molina,~A. Monte Carlo simulation of
  thermo-responsive charged nanogels in salt-free solutions. \emph{Soft Matter}
  \textbf{2013}, \emph{9}, 7086--7094\relax
\mciteBstWouldAddEndPuncttrue
\mciteSetBstMidEndSepPunct{\mcitedefaultmidpunct}
{\mcitedefaultendpunct}{\mcitedefaultseppunct}\relax
\EndOfBibitem
\bibitem[Jha \latin{et~al.}(2011)Jha, Zwanikken, Detcheverry, De~Pablo, and
  De~La~Cruz]{jha2011theoretical}
Jha,~P.~K.; Zwanikken,~J.~W.; Detcheverry,~F.~A.; De~Pablo,~J.~J.;
  De~La~Cruz,~M.~O. Study of volume phase transitions in polymeric nanogels by
  theoretically informed coarse-grained simulations. \emph{Soft Matter}
  \textbf{2011}, \emph{7}, 5965--5975\relax
\mciteBstWouldAddEndPuncttrue
\mciteSetBstMidEndSepPunct{\mcitedefaultmidpunct}
{\mcitedefaultendpunct}{\mcitedefaultseppunct}\relax
\EndOfBibitem
\bibitem[Schroeder \latin{et~al.}(2015)Schroeder, Rudov, Lyon, Richtering,
  Pich, and Potemkin]{schroeder}
Schroeder,~R.; Rudov,~A.~A.; Lyon,~L.~A.; Richtering,~W.; Pich,~A.;
  Potemkin,~I.~I. Electrostatic Interactions and Osmotic Pressure of
  Counterions Control the pH-Dependent Swelling and Collapse of Polyampholyte
  Microgels with Random Distribution of Ionizable Groups. \emph{Macromolecules}
  \textbf{2015}, \emph{48}, 5914--5927\relax
\mciteBstWouldAddEndPuncttrue
\mciteSetBstMidEndSepPunct{\mcitedefaultmidpunct}
{\mcitedefaultendpunct}{\mcitedefaultseppunct}\relax
\EndOfBibitem
\bibitem[Hofzumahaus \latin{et~al.}(2018)Hofzumahaus, Hebbeker, and
  Schneider]{MC_aachen}
Hofzumahaus,~C.; Hebbeker,~P.; Schneider,~S. Monte Carlo simulations of weak
  polyelectrolyte microgels: PH-dependence of conformation and ionization.
  \emph{Soft Matter} \textbf{2018}, \emph{14}\relax
\mciteBstWouldAddEndPuncttrue
\mciteSetBstMidEndSepPunct{\mcitedefaultmidpunct}
{\mcitedefaultendpunct}{\mcitedefaultseppunct}\relax
\EndOfBibitem
\bibitem[Sean \latin{et~al.}(2018)Sean, Landsgesell, and Holm]{reaxchain}
Sean,~D.; Landsgesell,~J.; Holm,~C. Computer Simulations of Static and
  Dynamical Properties of Weak Polyelectrolyte Nanogels in Salty Solutions.
  \emph{Gels} \textbf{2018}, \emph{4}, 2\relax
\mciteBstWouldAddEndPuncttrue
\mciteSetBstMidEndSepPunct{\mcitedefaultmidpunct}
{\mcitedefaultendpunct}{\mcitedefaultseppunct}\relax
\EndOfBibitem
\bibitem[Quesada-P{\'e}rez \latin{et~al.}(2018)Quesada-P{\'e}rez,
  Maroto-Centeno, Mart{\'\i}n-Molina, and Moncho-Jord{\'a}]{effective_int}
Quesada-P{\'e}rez,~M.; Maroto-Centeno,~J.~A.; Mart{\'\i}n-Molina,~A.;
  Moncho-Jord{\'a},~A. Direct determination of forces between charged nanogels
  through coarse-grained simulations. \emph{Physical Review E} \textbf{2018},
  \emph{97}, 042608\relax
\mciteBstWouldAddEndPuncttrue
\mciteSetBstMidEndSepPunct{\mcitedefaultmidpunct}
{\mcitedefaultendpunct}{\mcitedefaultseppunct}\relax
\EndOfBibitem
\bibitem[Gnan \latin{et~al.}(2017)Gnan, Rovigatti, Bergman, and
  Zaccarelli]{silicomicrogel}
Gnan,~N.; Rovigatti,~L.; Bergman,~M.; Zaccarelli,~E. In Silico Synthesis of
  Microgel Particles. \emph{Macromolecules} \textbf{2017}, \emph{50},
  8777--8786\relax
\mciteBstWouldAddEndPuncttrue
\mciteSetBstMidEndSepPunct{\mcitedefaultmidpunct}
{\mcitedefaultendpunct}{\mcitedefaultseppunct}\relax
\EndOfBibitem
\bibitem[Ninarello \latin{et~al.}(2019)Ninarello, Crassous, Paloli, Camerin,
  Gnan, Rovigatti, Schurtenberger, and Zaccarelli]{andrea2019preprint}
Ninarello,~A.; Crassous,~J.~J.; Paloli,~D.; Camerin,~F.; Gnan,~N.;
  Rovigatti,~L.; Schurtenberger,~P.; Zaccarelli,~E. Advanced modelling of
  microgel structure across the volume phase transition. \emph{arXiv preprint
  arXiv:1901.11495} \textbf{2019}, \relax
\mciteBstWouldAddEndPunctfalse
\mciteSetBstMidEndSepPunct{\mcitedefaultmidpunct}
{}{\mcitedefaultseppunct}\relax
\EndOfBibitem
\bibitem[Grest and Kremer(1986)Grest, and Kremer]{grest1986molecular}
Grest,~G.~S.; Kremer,~K. Molecular dynamics simulation for polymers in the
  presence of a heat bath. \emph{Physical Review A} \textbf{1986}, \emph{33},
  3628\relax
\mciteBstWouldAddEndPuncttrue
\mciteSetBstMidEndSepPunct{\mcitedefaultmidpunct}
{\mcitedefaultendpunct}{\mcitedefaultseppunct}\relax
\EndOfBibitem
\bibitem[Soddemann \latin{et~al.}(2001)Soddemann, D{\"u}nweg, and
  Kremer]{amphiphilic}
Soddemann,~T.; D{\"u}nweg,~B.; Kremer,~K. A generic computer model for
  amphiphilic systems. \emph{The European Physical Journal E} \textbf{2001},
  \emph{6}, 409--419\relax
\mciteBstWouldAddEndPuncttrue
\mciteSetBstMidEndSepPunct{\mcitedefaultmidpunct}
{\mcitedefaultendpunct}{\mcitedefaultseppunct}\relax
\EndOfBibitem
\bibitem[Hunter and White(1987)Hunter, and White]{hunter}
Hunter,~R.~J.; White,~L.~R. \emph{Foundations of colloid science}; Oxford
  science publications v. 1; Clarendon Press, 1987\relax
\mciteBstWouldAddEndPuncttrue
\mciteSetBstMidEndSepPunct{\mcitedefaultmidpunct}
{\mcitedefaultendpunct}{\mcitedefaultseppunct}\relax
\EndOfBibitem
\bibitem[Levin(2004)]{levin}
Levin,~Y. Introduction to statistical mechanics of charged systems.
  \emph{Brazilian Journal of Physics} \textbf{2004}, \emph{34},
  1158--1176\relax
\mciteBstWouldAddEndPuncttrue
\mciteSetBstMidEndSepPunct{\mcitedefaultmidpunct}
{\mcitedefaultendpunct}{\mcitedefaultseppunct}\relax
\EndOfBibitem
\bibitem[Lopez \latin{et~al.}(2019)Lopez, Scotti, Brugnoni, and
  Richtering]{kuhn_richtering}
Lopez,~C.~G.; Scotti,~A.; Brugnoni,~M.; Richtering,~W. The Swelling of Poly
  (Isopropylacrylamide) Near the $\theta$ Temperature: A Comparison between
  Linear and Cross-Linked Chains. \emph{Macromolecular Chemistry and Physics}
  \textbf{2019}, \emph{220}, 1800421\relax
\mciteBstWouldAddEndPuncttrue
\mciteSetBstMidEndSepPunct{\mcitedefaultmidpunct}
{\mcitedefaultendpunct}{\mcitedefaultseppunct}\relax
\EndOfBibitem
\bibitem[Landsgesell \latin{et~al.}(2019)Landsgesell, Nov{\'a}, Rud,
  Uhl{\'\i}k, Sean, Hebbeker, Holm, and Ko{\v{s}}ovan]{weak_ion_simulations}
Landsgesell,~J.; Nov{\'a},~L.; Rud,~O.; Uhl{\'\i}k,~F.; Sean,~D.; Hebbeker,~P.;
  Holm,~C.; Ko{\v{s}}ovan,~P. Simulations of ionization equilibria in weak
  polyelectrolyte solutions and gels. \emph{Soft matter} \textbf{2019},
  \emph{15}, 1155--1185\relax
\mciteBstWouldAddEndPuncttrue
\mciteSetBstMidEndSepPunct{\mcitedefaultmidpunct}
{\mcitedefaultendpunct}{\mcitedefaultseppunct}\relax
\EndOfBibitem
\bibitem[Tuckerman(2010)]{tuckerman}
Tuckerman,~M. \emph{Statistical mechanics: theory and molecular simulation};
  Oxford university press, 2010\relax
\mciteBstWouldAddEndPuncttrue
\mciteSetBstMidEndSepPunct{\mcitedefaultmidpunct}
{\mcitedefaultendpunct}{\mcitedefaultseppunct}\relax
\EndOfBibitem
\bibitem[Deserno and Holm(1998)Deserno, and Holm]{p3m}
Deserno,~M.; Holm,~C. How to mesh up Ewald sums. I. A theoretical and numerical
  comparison of various particle mesh routines. \emph{The Journal of chemical
  physics} \textbf{1998}, \emph{109}, 7678--7693\relax
\mciteBstWouldAddEndPuncttrue
\mciteSetBstMidEndSepPunct{\mcitedefaultmidpunct}
{\mcitedefaultendpunct}{\mcitedefaultseppunct}\relax
\EndOfBibitem
\bibitem[Plimpton(1995)]{LAMMPS}
Plimpton,~S. Fast Parallel Algorithms for Short-Range Molecular Dynamics.
  \emph{Journal of Computational Physics} \textbf{1995}, \emph{117},
  1--19\relax
\mciteBstWouldAddEndPuncttrue
\mciteSetBstMidEndSepPunct{\mcitedefaultmidpunct}
{\mcitedefaultendpunct}{\mcitedefaultseppunct}\relax
\EndOfBibitem
\bibitem[Stieger \latin{et~al.}(2004)Stieger, Richtering, Pedersen, and
  Lindner]{fuzzy}
Stieger,~M.; Richtering,~W.; Pedersen,~J.~S.; Lindner,~P. Small-angle neutron
  scattering study of structural changes in temperature sensitive microgel
  colloids. \emph{The Journal of chemical physics} \textbf{2004}, \emph{120},
  6197--6206\relax
\mciteBstWouldAddEndPuncttrue
\mciteSetBstMidEndSepPunct{\mcitedefaultmidpunct}
{\mcitedefaultendpunct}{\mcitedefaultseppunct}\relax
\EndOfBibitem
\bibitem[Conley \latin{et~al.}(2016)Conley, N{\"o}jd, Braibanti,
  Schurtenberger, and Scheffold]{superresolution}
Conley,~G.~M.; N{\"o}jd,~S.; Braibanti,~M.; Schurtenberger,~P.; Scheffold,~F.
  Superresolution microscopy of the volume phase transition of pNIPAM
  microgels. \emph{Colloids and Surfaces A: Physicochemical and Engineering
  Aspects} \textbf{2016}, \emph{499}, 18--23\relax
\mciteBstWouldAddEndPuncttrue
\mciteSetBstMidEndSepPunct{\mcitedefaultmidpunct}
{\mcitedefaultendpunct}{\mcitedefaultseppunct}\relax
\EndOfBibitem
\bibitem[Shibayama \latin{et~al.}(1992)Shibayama, Tanaka, and
  Han]{tanaka_ffactors}
Shibayama,~M.; Tanaka,~T.; Han,~C.~C. Small angle neutron scattering study on
  poly (N-isopropyl acrylamide) gels near their volume-phase transition
  temperature. \emph{The Journal of chemical physics} \textbf{1992}, \emph{97},
  6829--6841\relax
\mciteBstWouldAddEndPuncttrue
\mciteSetBstMidEndSepPunct{\mcitedefaultmidpunct}
{\mcitedefaultendpunct}{\mcitedefaultseppunct}\relax
\EndOfBibitem
\bibitem[Camerin \latin{et~al.}(2018)Camerin, Gnan, Rovigatti, and
  Zaccarelli]{camerin2018modelling}
Camerin,~F.; Gnan,~N.; Rovigatti,~L.; Zaccarelli,~E. Modelling realistic
  microgels in an explicit solvent. \emph{Scientific reports} \textbf{2018},
  \emph{8}, 14426\relax
\mciteBstWouldAddEndPuncttrue
\mciteSetBstMidEndSepPunct{\mcitedefaultmidpunct}
{\mcitedefaultendpunct}{\mcitedefaultseppunct}\relax
\EndOfBibitem
\bibitem[Kobayashi and Winkler(2016)Kobayashi, and
  Winkler]{conformational_prop_nanogels}
Kobayashi,~H.; Winkler,~R.~G. Universal conformational properties of polymers
  in ionic nanogels. \emph{Scientific Reports} \textbf{2016}, \emph{6},
  19836\relax
\mciteBstWouldAddEndPuncttrue
\mciteSetBstMidEndSepPunct{\mcitedefaultmidpunct}
{\mcitedefaultendpunct}{\mcitedefaultseppunct}\relax
\EndOfBibitem
\bibitem[Moreno and Lo~Verso(2018)Moreno, and Lo~Verso]{Moreno}
Moreno,~A.~J.; Lo~Verso,~F. Computational investigation of microgels: synthesis
  and effect of the microstructure on the deswelling behavior. \emph{Soft
  matter} \textbf{2018}, \emph{14}, 7083--7096\relax
\mciteBstWouldAddEndPuncttrue
\mciteSetBstMidEndSepPunct{\mcitedefaultmidpunct}
{\mcitedefaultendpunct}{\mcitedefaultseppunct}\relax
\EndOfBibitem
\bibitem[Li \latin{et~al.}(2007)Li, Zuo, Guo, Cai, Tang, and
  Yang]{li2007volume}
Li,~X.; Zuo,~J.; Guo,~Y.; Cai,~L.; Tang,~S.; Yang,~W. Volume phase transition
  temperature tuning and investigation of the swelling--deswelling oscillation
  of responsive microgels. \emph{Polymer International} \textbf{2007},
  \emph{56}, 968--975\relax
\mciteBstWouldAddEndPuncttrue
\mciteSetBstMidEndSepPunct{\mcitedefaultmidpunct}
{\mcitedefaultendpunct}{\mcitedefaultseppunct}\relax
\EndOfBibitem
\bibitem[Nigro \latin{et~al.}(2015)Nigro, Angelini, Bertoldo, Bruni,
  Castelvetro, Ricci, Rogers, and Ruzicka]{valentina_JCP}
Nigro,~V.; Angelini,~R.; Bertoldo,~M.; Bruni,~F.; Castelvetro,~V.;
  Ricci,~M.~A.; Rogers,~S.; Ruzicka,~B. Local structure of temperature and
  pH-sensitive colloidal microgels. \emph{The Journal of chemical physics}
  \textbf{2015}, \emph{143}, 114904\relax
\mciteBstWouldAddEndPuncttrue
\mciteSetBstMidEndSepPunct{\mcitedefaultmidpunct}
{\mcitedefaultendpunct}{\mcitedefaultseppunct}\relax
\EndOfBibitem
\bibitem[Bergman \latin{et~al.}(2019)Bergman, Pedersen, Schurtenberger, and
  Boon]{maxime_mie}
Bergman,~M.~J.; Pedersen,~J.~S.; Schurtenberger,~P.; Boon,~N. Morphologies of
  charge-regulating ionic microgels. \emph{in preparation} \textbf{2019},
  \relax
\mciteBstWouldAddEndPunctfalse
\mciteSetBstMidEndSepPunct{\mcitedefaultmidpunct}
{}{\mcitedefaultseppunct}\relax
\EndOfBibitem
\bibitem[Pe{\~n}a-Rodr{\'\i}guez \latin{et~al.}(2011)Pe{\~n}a-Rodr{\'\i}guez,
  Gonz{\'a}lez~P{\'e}rez, and Pal]{mie_lab}
Pe{\~n}a-Rodr{\'\i}guez,~O.; Gonz{\'a}lez~P{\'e}rez,~P.~P.; Pal,~U. MieLab: a
  software tool to perform calculations on the scattering of electromagnetic
  waves by multilayered spheres. \emph{International Journal of Spectroscopy}
  \textbf{2011}, \emph{2011}\relax
\mciteBstWouldAddEndPuncttrue
\mciteSetBstMidEndSepPunct{\mcitedefaultmidpunct}
{\mcitedefaultendpunct}{\mcitedefaultseppunct}\relax
\EndOfBibitem
\bibitem[Nigro \latin{et~al.}(2015)Nigro, Angelini, Bertoldo, Castelvetro,
  Ruocco, and Ruzicka]{nigro2015dynamic}
Nigro,~V.; Angelini,~R.; Bertoldo,~M.; Castelvetro,~V.; Ruocco,~G.; Ruzicka,~B.
  Dynamic light scattering study of temperature and pH sensitive colloidal
  microgels. \emph{Journal of Non-Crystalline Solids} \textbf{2015},
  \emph{407}, 361--366\relax
\mciteBstWouldAddEndPuncttrue
\mciteSetBstMidEndSepPunct{\mcitedefaultmidpunct}
{\mcitedefaultendpunct}{\mcitedefaultseppunct}\relax
\EndOfBibitem
\bibitem[Philippe \latin{et~al.}(2018)Philippe, Truzzolillo, Galvan-Myoshi,
  Dieudonn{\'e}-George, Trappe, Berthier, and Cipelletti]{philippe}
Philippe,~A.-M.; Truzzolillo,~D.; Galvan-Myoshi,~J.; Dieudonn{\'e}-George,~P.;
  Trappe,~V.; Berthier,~L.; Cipelletti,~L. Glass transition of soft colloids.
  \emph{Physical Review E} \textbf{2018}, \emph{97}, 040601\relax
\mciteBstWouldAddEndPuncttrue
\mciteSetBstMidEndSepPunct{\mcitedefaultmidpunct}
{\mcitedefaultendpunct}{\mcitedefaultseppunct}\relax
\EndOfBibitem
\end{mcitethebibliography}



\section*{Supplementary material}

# S1. Box size effects

It is a known fact that long-range correlations may arise in microgel systems, given the single particle intrinsically inhomogeneous structure in which monomer charges remain fixed, while counterions can freely move throughout the available particle volume. These long-range correlations in the counterion cloud surrounding the particles can strongly affect the microgel swelling properties. In particular, it has been shown that changes in counterions density profile modifies the osmotic pressure balance, causing the deswelling of microgels when the suspension concentration is raised. If this effect happens for neutral microgels in overlapping conditions,[1] for ionic microgel it plays a role already at packing fractions lower than the overlapping ones, though relatively high.[2] In addition, at low packing fractions and highly de-ionised conditions, the hydrodynamic radius has been proven to be strongly concentration-dependent, due to a change in the extension of the external chains, which significantly depends on the amount of the screening from the counterions, in turn determined by their concentration.[2]

Therefore, within our explicit counterions simulations, specific care must be taken to deal with long-range electrostatic interactions. Since we focused our investigation on the study of single particle properties at high dilution, we performed a preliminary analysis on the effects of the choice of the side $L$ of the (cubic) simulation box. Most importantly, we needed to identify a suitable value of $L$, which was large enough to avoid situations where periodic replicas would be able to feel each other, generating spurious effects. To this aim, we initially performed simulations of a smaller microgel with $N \sim 5000$ monomers with $f = 0.20$ and $c = 3.2\%$, assembled in a spherical cavity of radius $R_0 = 15\sigma$, in different box sizes with $40\sigma \leq L \leq 300\sigma$.

Looking at the variation of the total potential energy per bead $E/N$, shown in fig. S1, the long-range effects of electrostatic forces are evident, resulting in as a monotonic increase with $L$, mainly due to the decrease of average couterion concentration. The figure also shows the average bond energy, which remains unaffected by the change in $L$. Both those effects

are trivial and do not help choosing the correct box side.

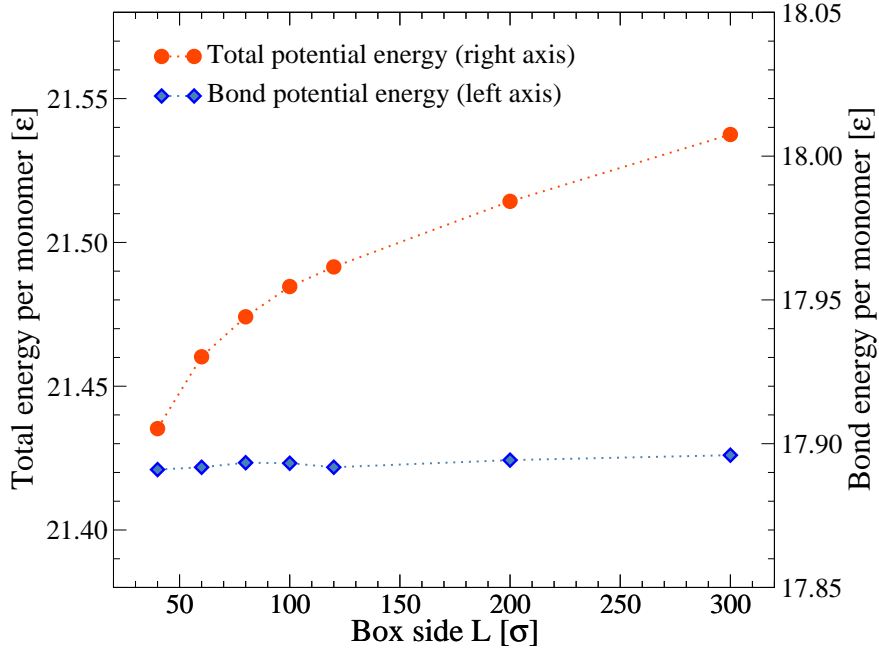

Figure S1: Total potential energy per monomer (scale on the left-axis) and bond potential energy per monomer (scale on the right axis) as a function of the side of the simulation box $L$. In the latter, only the contribution coming from $V_{FENE}$ is considered.

Relevant insights can be gained by looking at the evolution of the overall dimension and charge of the microgel particle when changing the box size. In Fig. S2 we report the dependence of the gyration radius on the box size. Since this observable mainly takes into account the contribution of the core of the particle, in order to also consider the extent of the corona, we also report the corresponding variation with $L$ of the upper limit of the hydrodynamic radius $R^*$ defined as the radius at which the spatial integral of the monomers density profile is equal to $N$.[3] Both $R_g$ and $R^*$ do not appreciably vary beyond $L \sim 80\sigma$, suggesting that the size of both the core and the corona of the microgel are stable for large boxes. Similarly, the charge contained inside a sphere of radius $R_g$, $Q(R_g)$, remains quite constant beyond $L \sim 80\sigma$, while the charge contained in a sphere of radius $R^*$, $Q(R^*)$, which also accounts for the counterions condensed on the corona of the particle, is found to increase quickly for small values of $L$ and more slowly at high values of $L$. On the basis of this last result, we select $L = 80\sigma$ as an optimal box size and we thus use the corresponding value

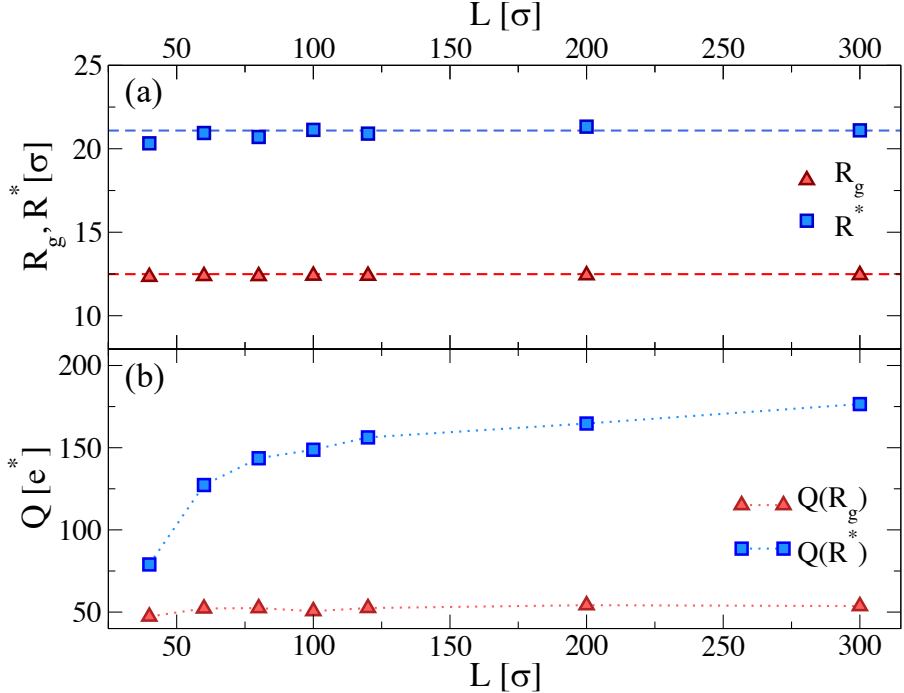

Figure S2: (a) Radius of gyration $R_g$ (triangles) and hydrodynamic radius $R^*$ (squares), defined as in the text, as a function of $L$; (b) charge contained in a sphere of radius $R_g$ (triangles) and of radius $R^*$ (squares), given in units of $e^*$, as a function of $L$.

$L = 300\sigma$ for the larger microgels ($N \approx 42000$) studied in the manuscript.

## S2. Choice of the counterion size

We also performed a preliminary analysis on the choice of the counterion size to assess its effects on the microgel properties across the volume phase transition. Clearly, a precise assessment of the coarse-graining size $\sigma_c$ of counterions, comprehensive of their hydration shell, is not a simple task. In reality, the value of $\sigma_c$ depends upon the ion species, the temperature, and the local environment wherein the ion is located (if the ion is confined within a dense polymer mesh the structure of its hydration shell could be different than in solution). However from *ab initio* molecular dynamics simulations it is found to be of the order of a few Å,[4] i.e. sensitively smaller than the Kuhn length of the polymers we are studying. Previous studies[3,5–7] had assigned to counterions the same excluded volume of

monomers. However, we found that such a choice, although having little effect on microgels in the swollen state, dramatically alters their internal structure and swelling properties at high $\alpha$ values. This is due to the fact that, when the solvophobic attraction increases and the microgel shrinks, excluded volume effects become important, so that the counterions become partially trapped in the interior of the microgels, giving rise to an oscillatory charge density profile.

When $\alpha$ increases the polymer beads tend to come in contact with each other in order to minimise the energy. This acts at the same time as the electrostatic attraction between charged monomers and counterions. At sufficiently high $\alpha$, the solvophobic attraction should win, but if the size of the counterions is too large, the additional steric repulsion will hinder the formation of bead-bead contacts and cause the presence of oscillations in the density profile of both monomers and counterions, as shown Fig. S3. We see that these effects are

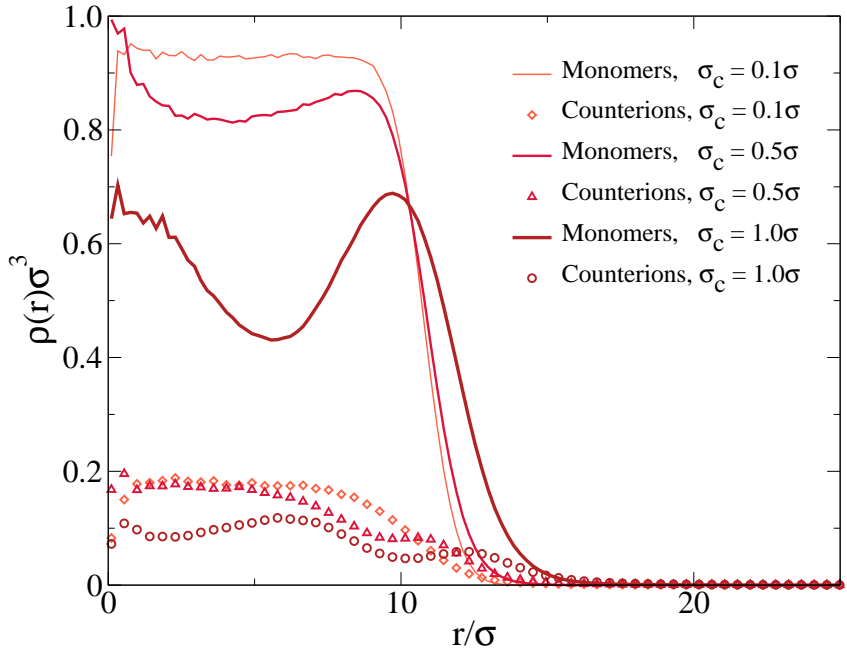

Figure S3: Density profiles of charged monomers (solid lines) and counterions (symbols) for the fully collapsed state ($\alpha = 1.50$) and different sizes of the counterions, i.e. $\sigma_c = 1.0, 0.5, 0.1\sigma$. Data are averaged over 4 different network topologies for microgels with $N \sim 5000$.

very strong for $\sigma_c = \sigma$ and still slightly present for $\sigma_c = 0.5\sigma$, leading to a decrease in the shrinking capability of the microgel at high $\alpha$ values. However, they disappear for $\sigma_c = 0.1\sigma$

5

that is the counterion size used throughout the manuscript.

## S3. Additional details on the form factors fits

In this section we show the form factors of the neutral and Debye-Hückel microgels with $f = 0.20$ at different values of $\alpha$ and we compare them with the fits obtained through the extended fuzzy sphere model with a modified Lorentzian (Eqns. 12 and 13 of the main text). We also comment the evolution of the fits parameters as a function of $\alpha$ and conclude discussing in more detail the capabilities of these models to describe the structure of charged microgels with explicit counterions.

As we can see in Fig. S4, all the data are well described by the model at all studied values of $\alpha$ for both types of microgels. In order to improve statistics, data have been averaged over 4 network topologies.

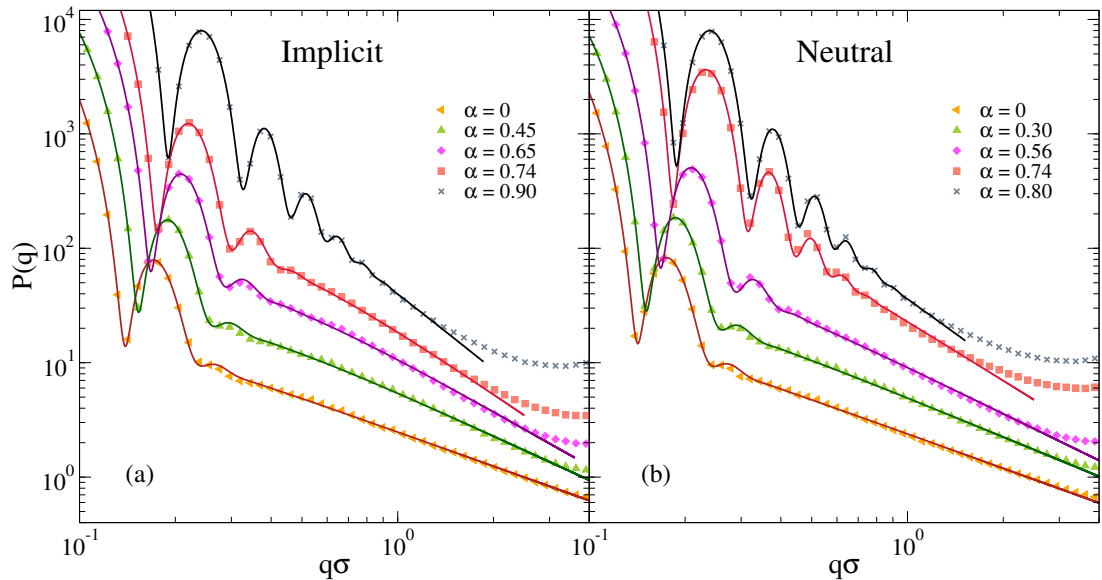

Figure S4: Zoom of the form factors of Figure 9(b,c) of the main text for (a) implicit (DH) and (b) neutral microgels at different values of $\alpha$. Symbols are simulation data, while lines are fits according to Eqns. 12 and 13 of the main text for $q\sigma \lesssim 3$. Data sets corresponding to the different $\alpha$ values have been vertically shifted for a matter of clarity.

Fig. S5 shows the evolution of the parameters obtained from the fits for the neutral model, the implicit one, and the explicit above the VPT. Panels S5(a,c) display the core radius $R$ and

the surface thickness $\sigma_{\text{surf}}$, characterising the structure at large length scales, or equivalently in the small-$q$ region of the Fourier space, modelled with and extended fuzzy sphere function (Eq. 12). They both decrease as a function of the effective temperature $\alpha$, because of the global shrinking of the microgels, as expected. Panels S5(b,d) show the evolution of the

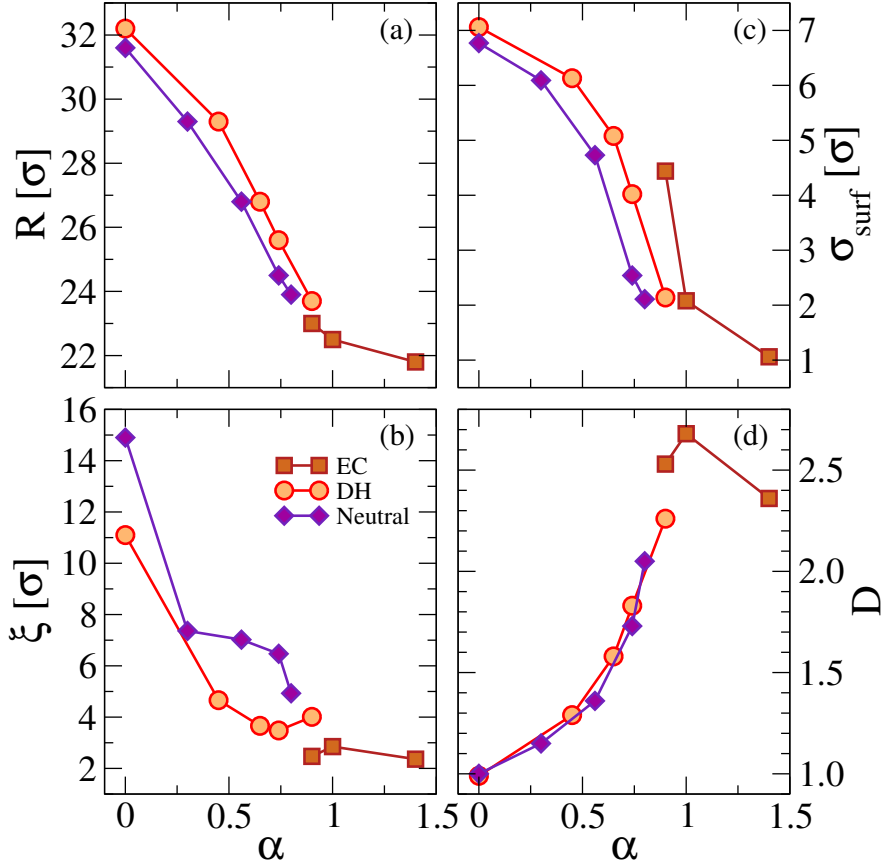

Figure S5: The four panels display the evolution of the fuzzy sphere parameters $R$ and $\sigma_{\text{surf}}$, the network correlation length $\xi$ and the fractal-dimension exponent $D$ as a function of $\alpha$, computed by fitting the form factors with Eqns. 12 and 13 of the main text. Three sets of data are shown in each panel: those relative to the microgel with explicit counterions for $\alpha > \alpha_{\text{VPT}}$ (squares), those relative to the Debye-Hückel model (circles), and those relative to the neutral case (diamonds).

correlation length $\xi$ and the fractal dimension exponent $D$ of the Lorentzian contribution (Eq. 13), which accounts for the polymer structure of the network, that is mostly visible in the intermediate- and high-$q$ region. As expected, $D$ increases when increasing $\alpha$. This indicates the coarsening of the network acted by the solvophobic attraction, which brings close to each other also monomers that are not chemically linked, enhancing the effective connectivity of

the network. The correlation length $\xi$ decreases with increasing temperature, suggesting that the average size of inhomogeneous regions in the polymer networks reasonably decreases when the microgel shrinks, becoming more and more compact and homogeneous. This seems to be also the case for the Debye-Hückel model, showing a qualitative agreement with experimental data,[8] apart from the largest $\alpha$ values, where the errors on these parameters become too large. Indeed the $q$-range wherein the Lorentzian contribution is found becomes smaller and smaller as $\alpha$ grows, being preempted by the growth of the peaks of the fuzzy sphere. Finally, for $q\sigma \gtrsim 2-3$ the data resolution is limited by the finite size of our numerical model.

When it comes to the microgels with explicit counterions, the modified shape of the corona and the absence of peaks beyond the first one are responsible for the fact that for small values of $\alpha$ a modification of the fitting model must be adopted. We thus only use the modified Lorentzian to fit separately the behavior of the regimes at intermediate $q$. The resulting fit parameters are summarised in fig. S6 and an extensive discussion of these results is reported in the main text. Here we further notice that finite size effects are responsible for increasing the error on the fit parameters when increasing $\alpha$. Particularly, for $\alpha = 0.74$ the first flat region of $P(q)$ extends over a too narrow $q$-interval to get meaningful results for the Lorentzian fit. Hence we constrain the fractal dimension to be close to that observed for smaller values of $\alpha$, i.e. $D \simeq 0.75$, because it seems to remain constant. This leads us to a value of $\xi$ that is in line with the other $\alpha$ values, that decreases with increasing $\alpha$. It is also larger for the first Lorentzian function than for the second, in agreement with the interpretation that the structure of the network in the corona (which dominates $P(q)$ at intermediate $q$ values) is more open than that in the core (which dominates $P(q)$ at larger $q$ values).

Above the VPT, the form factors are again well represented by an extended fuzzy sphere model plus a modified Lorentzian, as for the implicit and neutral microgels. These parameters are found in fig. S5. We notice that the Lorentzian parameters of the fit are affected by large errors and their value become very sensitive to the fitting range. This is due to the fact that

the region wherein the Lorentzian contribution dominates $P(q)$ shrinks with increasing $\alpha$.

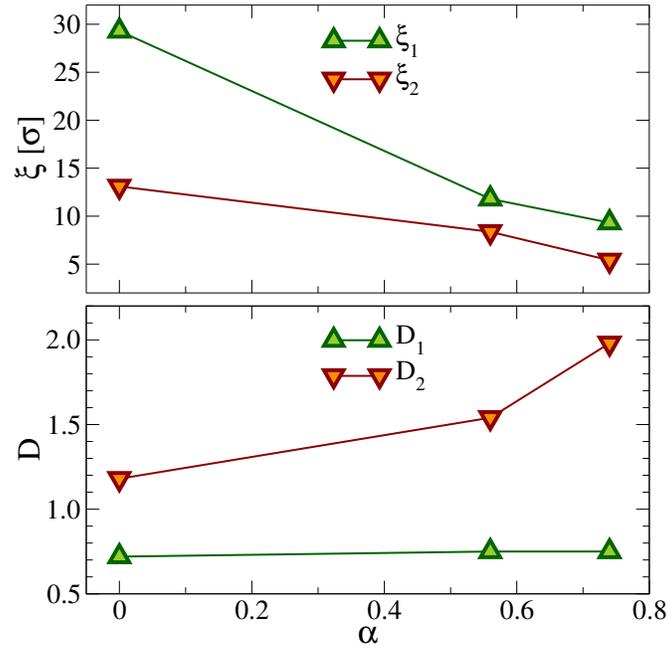

Figure S6: Evolution of the network correlation lengths $\xi_1$ and $\xi_2$ and of the fractal-dimension exponents $D_1$ and $D_2$ as a function of $\alpha$. The parameters have been extrapolated by fitting the two flat regimes of the form factors of microgels with explicit counterions with a modified Lorentzian function (Eq. 13 of the main text).

# References

(1) Mohanty, P. S.; Nöjd, S.; van Gruijthuijsen, K.; Crassous, J. J.; Obiols-Rabasa, M.; Schweins, R.; Stradner, A.; Schurtenberger, P. Interpenetration of polymeric microgels at ultrahigh densities. *Scientific Reports* **2017**, *7*, 1487.

(2) Nöjd, S.; Holmqvist, P.; Boon, N.; Obiols-Rabasa, M.; Mohanty, P. S.; Schweins, R.; Schurtenberger, P. Deswelling behaviour of ionic microgel particles from low to ultrahigh densities. *Soft matter* **2018**, *14*, 4150–4159.

(3) Quesada-Pérez, M.; Maroto-Centeno, J. A.; Martín-Molina, A.; Moncho-Jordá, A. Direct determination of forces between charged nanogels through coarse-grained simulations. *Physical Review E* **2018**, *97*, 042608.

(4) Tuckerman, M.; Laasonen, K.; Sprik, M.; Parrinello, M. Ab initio molecular dynamics simulation of the solvation and transport of hydronium and hydroxyl ions in water. *The Journal of chemical physics* **1995**, *103*, 150–161.

(5) Kobayashi, H.; Halver, R.; Sutmann, G.; Winkler, R. G. Polymer conformations in ionic microgels in the presence of salt: Theoretical and mesoscale simulation results. *Polymers* **2017**, *9*, 15.

(6) Hofzumahaus, C.; Hebbeker, P.; Schneider, S. Monte Carlo simulations of weak polyelectrolyte microgels: PH-dependence of conformation and ionization. *Soft Matter* **2018**, *14*.

(7) Sean, D.; Landsgesell, J.; Holm, C. Computer Simulations of Static and Dynamical Properties of Weak Polyelectrolyte Nanogels in Salty Solutions. *Gels* **2018**, *4*, 2.

(8) Nigro, V.; Angelini, R.; Bertoldo, M.; Bruni, F.; Ricci, M. A.; Ruzicka, B. Dynamical behavior of microgels of interpenetrated polymer networks. *Soft matter* **2017**, *13*, 5185–5193.



\end{document}